\documentclass[MSNbibl,nameyear,dvips]{arxstspdf}
\usepackage{flushend}
\usepackage{stfloats}
\usepackage{graphicx}
\volume{29}
\issue{1}
\pubyear{2014}
\firstpage{144}
\lastpage{163}
\doi{10.1214/13-STS449} 

\makeatletter

\def\cal{\mathcal}

\renewcommand{\citep}[1]{(\citeauthor{#1}, \citeyear{#1})}
\newcommand{\eqref}[1]{(\ref{#1})}

\makeatother

\begin{document}
\begin{frontmatter}

\title{Selecting a Biased-Coin Design}
\runtitle{Selecting a Biased-Coin Design}

\begin{aug}
\author[a]{\fnms{Anthony C.} \snm{Atkinson}\corref{}\ead[label=e1]{a.c.atkinson@lse.ac.uk}}
\runauthor{A. C. Atkinson}

\affiliation{London School of Economics}

\address[a]{Anthony C. Atkinson is Emeritus Professor, Department of Statistics, London School of Economics, London WC2A 2AE, United Kingdom.}

\end{aug}

%
\begin{abstract}
Biased-coin designs are used in clinical trials to allocate treatments
with some randomness while maintaining approximately equal allocation.
More recent rules are compared with Efron's
[\textit{Biometrika} \textbf{58}  (1971) 403--417] biased-coin rule and
extended to allow balance over covariates. The main properties are loss
of information, due to imbalance, and selection bias. Theoretical
results, mostly large sample, are assembled and assessed by
small-sample simulations. The properties of the rules fall into three
clear categories. A Bayesian rule is shown to have appealing
properties; at the cost of slight imbalance, bias is virtually
eliminated for large samples.
\end{abstract}

%
\begin{keyword}
\kwd{Clinical trial}
\kwd{covariate balancing}
\kwd{loss of information}
\kwd{optimum experimental design}
\kwd{random allocation}
\kwd{selection bias}
\end{keyword}\vspace*{6pt}

\end{frontmatter}
%
\section{Introduction}
\label{introsec}

It is now over forty years since \citet{efr71} introduced a
biased-coin design for the partially randomized sequential allocation
of one of two treatments. The intention was to provide approximate
balance whenever the experiment was stopped while providing
randomization to reduce biases. This was achieved by allocating the
under-represented treatment with a constant probability; Efron
preferred $ p = 2/3$. In the case of equal cumulative allocation to the
two treatments, allocation was made at random. Since that time there
have been many developments, recently reviewed by \citet
{biswasbhattacharya11} and compared by \citet{zhao2011}. In the
comparison of these designs the emphasis has tended to be on balance.
See, for example, \citet{baldi2008}. One purpose of this review is to
stress the importance of looking at both bias and balance in the
assessment of designs.

Efron's rule and its extensions are used when either there are no
prognostic factors or they are ignored by the design. It is now also
over thirty years since \citet{aca82} introduced a randomized version
of the sequential construction of optimum designs that can include
discrete or continuous prognostic factors and can be used for a wide
variety of error distributions. (A connection with Efron's rule was a
visit to Imperial College in 1981 by Rupert Miller and a discussion of
\citeauthor{brad80}, \citeyear{brad80}.) For normal homoscedastic models without covariates
this randomized rule can provide a series of alternatives to Efron's
rule with controllable randomization \citep{rls84b}. Consideration of
bias and loss as functions of the number of trials leads to the
division of the rules, both with and without covariates, into three
groups with very different properties. Balance is measured as an
effective loss in the number of patients due to imbalance. Plots of
loss against bias provide a cogent way to summarise the properties of
designs and to determine whether designs are admissible. Efron's
original proposal is not.

The paper starts with rules without covariates. Several such rules are
described in Section~\ref{rulesec}. Efron's rule can be extended by
taking values of $p$ other than $2/3$. The distribution of $D_n$, the
difference in the number of allocations to the two treatments, forms a
Markov chain, the steady-state properties of which were studied by
\citet{efr71}.\vadjust{\goodbreak} Exact distributional results for the chain are given by
\citet{markbill2010}. However, the design can become appreciably
unbalanced; the rules of \citet{soareswu83} and of \citet{ypchen99}
limit the maximum value of $D_n$. In addition to these rules, \citet
{baldis2004} introduced a family of ``adjustable'' biased-coin rules
that force balance more strongly as the difference in allocations
increases. Two rather different families of rules, derived from
considerations of the optimum design of experiments, are also
introduced in Section~\ref{rulesec}. The rules of \citeauthor{rls84a} (\citeyear
{rls84a,rls84b}) are a generalization of those of \citet{aca82} which
used randomized versions of the sequential construction of optimum
designs to provide allocation rules giving balance over covariates.
(Consideration of rules for covariate balance starts in Section~\ref{covarsec}.) The final family of rules are derived from the Bayesian
work of \citet{ball93} which explicitly balances inference and randomization.

The criteria for comparison of the rules, loss and selection bias, are
introduced in Section~\ref{asesssec}. It is assumed that the property
of interest is precise estimation of the treatment difference.
Section~\ref{ananumsec} provides a survey and comparison of results for
the bias and loss of the rules. Simulation is used to explore the
reliability of asymptotic results.

For Efron's biased coin, Section~\ref{efronanalsec}, the values of loss
from the steady-state distribution of $D_n$ are compared with
simulation results, showing good agreement for $p = 2/3$ and $n > 50$.
In Section~\ref{baldananusec} a \mbox{simple} approximation is found to the
steady-state distribution of $D_n$ for the adjustable-biased coin of
\citet{baldis2004}. This shows that the design strongly forces
balance and that the loss is small. However, for odd $n$, the bias is
the highest of those for the rules considered, a feature that is not
obvious from the original paper. For Smith's rule, Section~\ref{smithananumsec}, the asymptotic results are again compared with the
results of simulation.

The results show very clearly the importance of assessing rules both
for odd and even $n$. This point was unfortunately overlooked by \citet
{zhao2011}, all of whose extensive simulation results are for even
$n$. In Section~\ref{adjavsec} the averages of adjacent values of loss
and bias are used in the comparison of rules. Such averages remove the
effect of the parity of $n$ and confirm the general principle, for
example, \citet{aca2002}, that appreciably random rules have high loss
and low bias, whereas rules that force balance will have low loss but
high bias since forcing balance makes it easier to guess correctly
which treatment will be allocated next.

The admissibility plots, introduced by \citet{aca2002}, are given in
Section~\ref{admisssec} for the nine rules considered. These show the
very different behaviour of the three groups of rules. In particular,
Efron's coin is inadmissible compared with an instance of the rule of
\citet{baldis2004}. The new Bayesian rule, that initially forces
balance, becomes closer to random allocation as the sample size
increases, thus reducing selection bias for larger samples.

These biased-coin rules can be extended to include covariates through
balance over individual strata or by balance over the variables in a
linear model. Section~\ref{covarsec} introduces these approaches,
together with that of randomized versions of the sequential
construction of optimum designs for a linear model. The admissibility
of these allocation rules is evaluated in Section~\ref{normalsec}.
There is much common structure between the admissibility plots for
rules with covariates in Figure~\ref{coinf20} and those for rules
without covariates in Figure~\ref{coinf12}.

The purposes of the paper include the gathering together in one place
of theoretical results on the properties of the rules, which are
compared to small sample results by simulation. Another purpose is to
include the recent rule of \citet{baldis2004} in this framework and
to provide some tractable theoretical results on the properties of the
rule. A difference between rules without covariates and those with
continuous covariates is their strong dependence on the parity of $n$.
The paper closes in Section~\ref{concsec} with a discussion of
extensions to several treatments and to models other than homoscedastic
regression. We commend the Bayesian rule of Sections~\ref{bayesananumsec} and \ref{covarsec}. In the absence of covariates the
average loss for this procedure for large $n$ is equal to the loss of
information on one patient, a~small price to pay for the avoidance of
selection bias in an asymptotically efficient rule.

\section{Rules without Covariates}
\label{rulesec}

There are two treatments and $n$ patients of whom $N_i$ have received
treatment $T_i\  (i=1,2)$. Because treatment allocation involves some
randomness, the $N_i$ are random variables. A further important
variable is the difference in the number of allocations of the two
treatments $D_n = N_1 - N_2$. The probability that patient $n+1$ is
assigned to treatment 1 is given by a function $F(n_1,n_2)$, $N_i = n_i$
$(i = 1,2).$

\subsection{Efron's Biased-Coin Design: Rule~E}

\label{efrsec}

In Efron's biased-coin design the allocations depend on $n_1$ and $n_2$
through the difference $D_n$:
\[
F_E(x) = \cases{ %
p, & $x<0,$
\vspace*{2pt}\cr
0.5, & $x= 0,$
\vspace*{2pt}\cr
q = 1-p, & $x>0,$}
\]
for $0.5 \le p \le1$ and $x = D_n$. Efron (Section~2) favours $p =
2/3$, which can easily be implemented using a six-sided die. For $ p =
0.5$ treatment allocation becomes random, being decided by tossing a
fair coin. There is no control over balance, but consistent successful
guessing of the next allocation is impossible.

For $p=1$ the rule becomes that of sequential design construction and
$|D_N|$ is either zero, when allocation is at random, or 1, when the
under-represented treatment is allocated. This deterministic rule will
be called \textit{Rule~D} and random allocation \textit{Rule~R}. These two
rules are at the extremes of all sensible rules---theoretically the
treatments could be allocated to increase imbalance, but not in a
practical context. Allocation with Rule~R does not depend on whether
$n$ is even or odd. But, for Rule~D the two values produce allocations
with extreme properties, random or deterministic. The variation of
properties with the parity of $n$ is a strong feature of Rule~E and,
particularly, of the rule in the next subsection.

\subsection{The Adjustable Biased-Coin Design: Rule~J}
\label{abcdblsec}

The correction toward balance in Rule~E depends only on the sign of
$D_n$ but not on its magnitude. The design may therefore sometimes
become appreciably unbalanced and several rules have been suggested to
reduce the variability of $D_n$. Part of the argument of this review is
that such concerns may be overstated, given the small effect of
appreciable imbalance on the statistical performance of the design for
moderate $n$.

\citet{baldis2004} introduced a rule in which the corrective force
increases with $|D_n|$:
%
\begin{equation}
F_{BA}(x) = \cases{ %
\displaystyle\frac{|x|^a}{1+ |x|^a}, &
$x<0,$
\vspace*{2pt}\cr
0.5, & $x= 0,$
\vspace*{2pt}\cr
\displaystyle\frac{1}{1+ |x|^a}, & $x>0,$} \label{baldirule}
\end{equation}
for $a \ge0$. \citeauthor{baldis2004} refer to their rule as ABCD.
However, all these letters have been used by \citet{aca2002}. To avoid
confusion, in the present paper the rule is called J for ``Ad\textit{j}ustable.''

In this rule a difference of one between treatments is treated as if it
were zero and the next treatment is allocated at random. The corrective
force increases with $|x|$. The value $a = 0$ gives Rule~R, whereas as
$a \rightarrow\infty$, the rule tends to Rule~D. \citet{baldis2004}
tabulate properties for $a$ from 1 to 4.

\subsection{Imbalance Tolerance}
\label{imtolsec}

In Rule~J the distribution of $D_n$ has support on $(-n, n)$, although,
as Tables~1--3 of \citet{baldis2004} show, for $a$ in the range 1--4,
the distribution is concentrated on a few values of $D_n$ near zero.
Two earlier rules were formulated to restrict the range of values of
$D_n$. In the balanced-coin design with imbalance tolerance of \citet
{ypchen99}, the rule is that of Efron except that there is a
reflecting barrier in the stochastic process for $D_n$ at~$\pm
b$:\looseness=1
\[
F_{IT}(x) = \cases{ %
1, & $x = -b,$
\vspace*{2pt}\cr
p, & $x<0,$
\vspace*{2pt}\cr
0.5, &$x= 0,$
\vspace*{2pt}\cr
q = 1-p, & $x>0,$
\vspace*{2pt}\cr
0, & $x = b.$}
\]\looseness=0
The ``big stick'' design of \citet{soareswu83} is obtained when $p =
1/2$. We do not investigate the properties of these rules, which have
an evolution of properties with $n$ that is similar to that of Rules~E
and J. These rules are related to the tractable approximation to the
properties of Rule~J given in Table~\ref{cointab1}.

\subsection{Rule~P: Permuted Block Design}


Deterministic allocation, Rule~D, can be thought of as allocating
conceptual blocks of length 2, ensuring balance
whenever $n$ is even. The blocks are ``conceptual'' since they are
not like the blocks in a conventional experiment; they do not
correspond to groups of units with common properties and they are
not included in the analysis. An extension is to allocate larger
randomized sequences, for example, AABABABB, ensuring balance when $n$
is a
multiple of eight, but not otherwise. \citet{efr71} explores some
properties of designs with block sizes up to 32; \citeauthor{rosl2002} (\citeyear{rosl2002}) in their
Figure~6.3 only go up to block size 10.

\subsection{Smith: Rule~S}
\label{smithsec}

\citet{rls84b} investigated a family of rules in which
%
\begin{equation}
F_S(n_1,n_2) = \frac{n_2^{\rho}}{n_1^{\rho} +
n_2^{\rho}}
\label{ruleS}
\end{equation}
for $\rho\ge0.$ As $\rho\rightarrow0$ we again obtain random
allocation and, as $\rho\rightarrow\infty$, Rule~D. Although the
allocation probabilities depend on both $n_1$ and $n_2$, \eqref{ruleS}
can be rewritten to show that the dependence on earlier allocations is
only through the ratio $D_n/n$. For the biased-coin rules described
above, dependence is directly on the difference $D_n$. The results of
Section~\ref{smithananumsec} show the effect of this distinction.

The family of rules was suggested by the designs of \citet{aca82} for
randomized allocation when there are covariates over which balance is
required, described in Section~\ref{covarsec}. In the absence of
covariates the model contains just two parameters, the mean treatment
effects $\mu_1$ and $\mu_2$. The D-optimum design maximizes the
determinant of the information matrix for the two parameters, in this
case minimizing the product of the variances, leading to a value of one
for $\rho$. Rule~A of Section~\ref{covarsec} (the D$_A$-optimum
design), minimizing $\operatorname{Var}(\hat{\mu}_1 - \hat{\mu}_2)$, is obtained when
$\rho= 2$.

It is simple to show that this rule, when $\rho= 1,$ is the same as
the adaptive biased-coin design of \citet{wei78}, who suggests the
linear rule
\[
F_W(n_1,n_2) = (1 - D_n/n)/2.
\]
Some of the results of Section~\ref{smithananumsec} therefore also
apply to Wei's rule.

\subsection{Bayesian Procedure: Rule~B}

\label{bayessec}

Both \citet{rls84b} and \citet{markbill2010} suggest that the rule
should be designed with statistical principles in mind, such as
variance of estimation and bias.

\citeauthor{rls84b} (\citeyear{rls84b}) in his Section~4 derives an expression for the mean squared
error of prediction when both variance and selection bias are included.
His conclusion is that $\rho$ should tend to zero as $n$ becomes large.
That is, for large $n$, the rule should become increasingly like random
allocation. This behaviour is achieved by the Bayesian rule of this section.

In his development of methods of comparison of designs with prognostic
factors, \citet{aca2002} introduces a Bayesian rule, derived from a
general approach of \citet{ball93}, which balances randomness and
precision of estimation through inclusion of a parameter $\gamma$. For
the allocation of one of two treatments in the absence of prognostic
factors, the rule is
%
\begin{eqnarray}\label{bayesrule}
&&F_B(n_1,n_2)
\nonumber
\\[-8pt]
\\[-8pt]
\nonumber
&&\qquad = \frac{\{1 + n_2/(nn_1)\}^{1/\gamma}} {
\{1 + n_2/(nn_1)\}^{1/\gamma} + \{1 + n_1/(nn_2)\}^{1/\gamma}}
\end{eqnarray}
$(0 \le\gamma\le1)$. As $n \rightarrow\infty$ with $\gamma= 1$
random allocation, Rule~R is obtained, whereas $\gamma\rightarrow0$
gives Rule~D for small $n$, that is, sequential construction of the
D$_A$-optimum design. In this case randomization is ignored.

The important feature that distinguishes this rule from the others is
its behaviour as a function of $n$. Initially, as the results in
Section~\ref{admisssec} show, when $n$ is small and $\gamma$ is also
small, 0.01 in the numerical example, the rule forces balance, behaving
like Rule~D. However, as $n$ increases, \eqref{bayesrule} shows that
the effect of imbalance on the allocation probability decreases. For
large $n$ the rule indeed behaves increasingly like random allocation.

\section{Assessing Rules: Bias and Loss}
\label{asesssec}

It is usual to assume that the observations are, at least
approximately, normally distributed with constant variance $\sigma^2$
and that the treatment difference $\mu_1 - \mu_2$ is the parameter of
interest. The effect of imbalance due to randomization is slightly to
increase the variance of the estimated difference
%
\begin{eqnarray}\label{vardiff}
\operatorname{Var} (\hat{\mu}_1 - \hat{\mu}_2)& =&
\sigma^2(1/n_1 + 1/n_2)
\nonumber
\\[-8pt]
\\[-8pt]
\nonumber
& =& n
\sigma^2/(n_1n_2).
\end{eqnarray}
For the balanced design with $n$ even,
%
\begin{equation}
\operatorname{Var}^* (\hat{\mu}_1 - \hat{\mu}_2) = 4\sigma
^2/n \quad (n = 2m), \label{vardiff2}
\end{equation}
where $m$ is an integer and $^*$ indicates ``for the optimum design.'' Of
course, there will always be an imbalance of one in the optimum design
when $n$ is odd ($D_{2m+1} = \pm1$), when
%
\begin{equation}
\hspace*{4pt}\quad\operatorname{Var}^* (\hat{\mu}_1 - \hat{\mu}_2) = 4\sigma
^2/(n - 1/n)\quad (n = 2m+1). \label{vardiff3}
\end{equation}
The difference between \eqref{vardiff2} and \eqref{vardiff3} when $n =
11$ is less than 1\% and decreases like $n^{-2}$. The distinction in
parity of $n$ will be ignored in variance comparisons.

\citet{bur96} suggested rewriting \eqref{vardiff} as
\[
\operatorname{Var} (\hat{\mu}_1 - \hat{\mu}_2) =
\frac{4\sigma^2}{n - L_n},
\]
where $L_n$ is defined to be the ``loss,'' that is, the effective number
of patients on whom information is lost due to the imbalance of the
design. Then, from \eqref{vardiff},
%
\begin{equation}
L_n = D^2_n/n \label{evenloss}
\end{equation}
for even $n$. For $n$ odd we could take $L_n = (D^2_n-1)/n$.
The value of $L_n$ is random, depending on the outcome of the
particular randomization. The expected value of $L_n$ is written
%
\begin{equation}
{\cal L}_n = (\operatorname{Var} D_n)/n, \label{expecloss}
\end{equation}
since, for the designs considered here, E$ D_n =0$.

An important related statistical quantity is the efficiency of the design,
%
\begin{equation}
E_n = \frac{\operatorname{Var}^*  (\hat{\mu}_1 - \hat{\mu
}_2)}{\operatorname{Var}  (\hat{\mu}_1 - \hat{\mu}_2)} = \frac{n - L_n}{n} = 1 -
L_n/n. \label{nocoveffic}
\end{equation}
These quantities are highly informative about the properties of all
designs considered.

The loss $L_n$ depends on the particular sequence of randomized
allocations. Interest here is in the expectation E$ L_n = {\cal L}_n$,
approximated by $\bar{L}_n$, the average over $n_{\mathrm{sim}}$
simulations. For Rule~S the expected value of the loss ${\cal L}_n $
has a constant limit as $n \rightarrow\infty$ (Section~\ref{smithananumsec}). For random allocation, a result that goes back at
least to \citet{cox51} is that ${\cal L}_{\infty} = 1.$ For the
biased-coin Rules~E and J, $\operatorname{Var} D_n$ approaches from below a finite
limit as $n \rightarrow\infty$. The loss therefore decreases with $n$.
Simulation results on the distribution of $L_n$ for rules with
covariates are presented by \citet{aca2003a}.

Statistical power is extremely important in the practical assessment of
designs for clinical trials. Plots like those in Figure~3.1 of \citet
{rosl2002}, and the similar plot in \citet{pocock83}, show how very
large the value of $D_n$ has to be to cause a measurable effect on
power. For both forms of behaviour of loss with increasing $n$, it
follows from \eqref{nocoveffic} that the efficiency of the design goes
to one as $n \rightarrow\infty$. Except for very small trials, the
average effect of imbalance on \eqref{vardiff} will be negligible. For
a particular randomization, the interpretation of $L_n$ as a number of
patients on whom information is lost leads directly to the calculation
of loss in power due to the imbalance from randomization.

Randomization and balance are in conflict. A numerical measure for
randomization is selection bias \citep{bh57} which measures the
ability to guess the next treatment to be allocated. Bias depends on
the design, the guessing strategy and, for some rules, the value of
$n$. For a particular combination of strategy and design the expected
bias ${\cal B}_n$ is estimated from $n_{\mathrm{sim}}$ simulations as
%
\begin{eqnarray}
\label{biasn} \bar{B}_n & = & (\mbox{number of correct guesses}\nonumber\\
&&\hspace*{4pt}\mbox{of
allocation to patient } n
\\
& &{} - \mbox{number of incorrect guesses})/n_{\mathrm{sim}}.\nonumber
\end{eqnarray}
This definition is similar to that of (4.2) of \citet{rls84b}.
The guessing strategy used in the numerical comparisons of the next
sections is the sensible one of guessing that the treatment for which
the allocation probability $p \ge0.5$ will be selected.

\citet{aca2002} calculated $\bar{B}_n$ from the binary variables in
\eqref{biasn}. It is, however, more efficient, as is done here, to
follow \citet{heritier2005} and average the expectations $2p -1$.

The customary justifications for randomization in experiments include
the avoidance of bias. Amongst many others, \citet{efr71} and \citet
{rls84b} consider that selection bias should not be an issue in
double-blind trials with treatment allocation made remotely from the
trial, although it may be if there are local attempts toward
institutional balance \citep{lagakos1984}. However, a trial without
randomization appears to lack objectivity. Accordingly, they study the
effect of biased-coin designs on freedom from accidental bias due to
omitted factors, including time trends and, in the case of \citet
{rls84b}, correlated errors and outliers. The conclusion of \citet
{rls84b} is that biased-coin designs that are not as random as Rule~R
provide good protection against several sources of bias and that
selection bias is a good measure of the properties of the design.

There are several related measures of selection bias. For example,
\citet{efr71} and \citet{markbill2010} use excess selection bias,
that is, the expected number of correct guesses in excess of those
expected when allocation is at random. A~slight advantage of ${\cal
B}_n$ is that the values lie between 0 and 1. However, one measure is a
linear function of the other, so that the ranking of rules is not
changed by the choice of measure.

\citeauthor{rls84b} (\citeyear{rls84b}) in his equation~4.2 defined the selection bias ${\cal B}_n$,
but found the average bias over all allocations up to $n$. Average bias
was also used by \citet{baldis2004} and by \citet{zhao2011}, who,
however, calculate the mean number of correct guesses. In the present
paper the measure used is $\bar{B}_n$, which refers solely to guesses
of the $n$th allocation. An advantage of this measure is that it
reveals whether there is a strong dependence of the bias on the parity
of $n$. A book length discussion of forms of selection bias is due to
\citet{berg2005}. \citet{prosch2012} briefly discuss some situations
in which blinding is impossible and selection bias may seriously distort
inferences.

\section{Analytical and Numerical Results on Bias and Loss}
\label{ananumsec}

Analytical results for bias and loss are available for some rules. They
are summarized by allocation rule in the following section. For Rules~E
and J the distribution of the values of $D_n$ forms a Markov chain. The
analytical results come from the steady-state distribution of $D_n$.
The value of ${\cal B}_n$ depends on the distribution of $D_{n-1}$,
whereas that of ${\cal L}_n$ requires the distribution of $D_n$. For
Rule~S the results are asymptotic and do not depend on the parity of
$n$. Simulation is used to check the properties of the rules for finite
$n$ and to show some properties of Rule~B.

There are also related results on the distribution of the number of
patients $N_i$ allocated to each treatment. For Rules E and J the
limiting Markov chain for the distribution of $D_n$ leads to the result
that $n^{-1/2}D_n$ goes to 0 in probability, so that $N_i$ goes to one
half. For Rule~S the proportions have an asymptotically normal
distribution depending on the value of $\rho$ in \eqref{ruleS}. The
details are in Section~\ref{smithananumsec}.

\begin{figure*}

\includegraphics{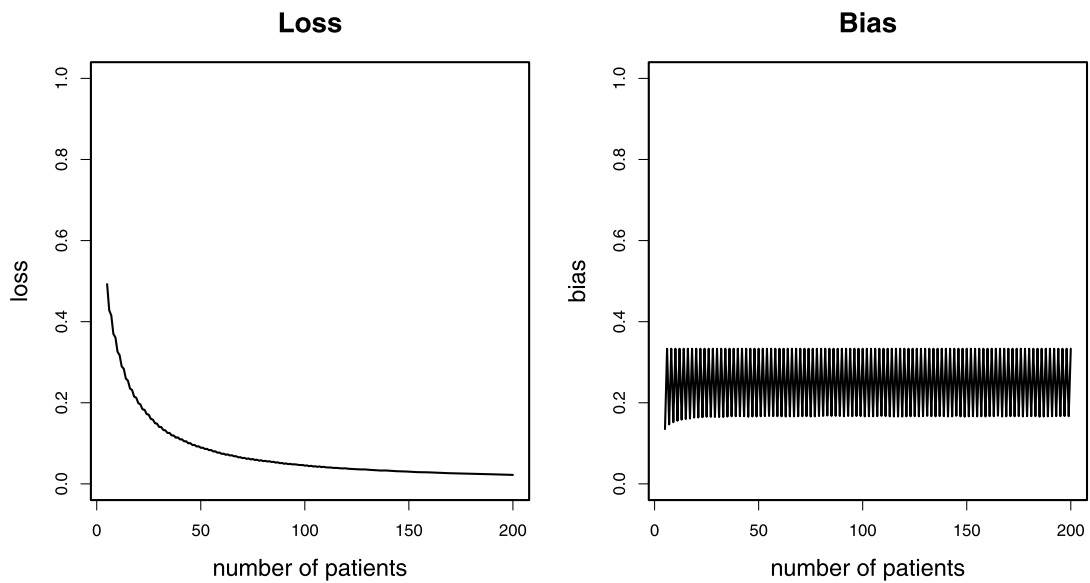}

\caption{Efron's coin with $p=2/3$---Rule E$(2/3)$. Left-hand panel,
average loss $\bar{L}_n$, right-hand panel, average bias $\bar{B}_n$.
100,000 simulations.}\label{coinf1}
\end{figure*}

\subsection{Efron's Biased Coin: Rule~E}
\label{efronanalsec}

If $D_{n-1} = 0$, allocation to patient $n$ is at random and the bias
is zero. For all other values the bias, conditional on the value of
$D_{n-1}$, is $2p-1$. The steady-state probabilities that $D_n = 0$ are
given by \citet{efr71} and by \citet{markbill2010} who both write $r
= p/q = p/(1-p).$ Then, for integer $m$,
\begin{eqnarray}\label{zeroprob}
\mathop{\lim}_{n \rightarrow\infty} P(D_{2m} = 0) = \frac{r-1}{r} =
\frac{2p-1}{p},
\nonumber
\\[-8pt]
\\[-8pt]
\eqntext{\mbox{with } P(D_{2m-1} = 0) = 0.}
\end{eqnarray}
When $n$ is even, that is, $n = 2m$, it follows from \eqref{zeroprob} that
\[
{\cal B}_{2m} = 2p - 1.
\]
When $n$ is odd, $P(D_{2m} \ne0) = (1-p)/p$ and
\[
{\cal B}_{2m-1} = (2p - 1) (1-p)/p.
\]
For the original coin with $p = 2/3$,
%
\begin{equation}\quad
{\cal B}_{2m}(2/3) = 1/3 \quad\mbox{and}\quad {\cal B}_{2m-1}(2/3) =
1/6. \label{efrbias}
\end{equation}

The right-hand panel of Figure~\ref{coinf1} shows average simulated
values $\bar{B}_n$ for Rule~E($2/3$), that is, Efron's coin. Although the
two values in \eqref{efrbias} are calculated from the steady-state
distribution of $D_n$, the figure shows that the bias immediately
settles down to oscillate between these two values.

For random allocation $p = 0.5$ and ${\cal B}_{n}(0.5) = 0$, whether
$n$ is odd or even. As $p$ increases, ${\cal B}_{2m} \rightarrow1$,
whereas ${\cal B}_{2m-1}$ has a maximum at the famous number $(\sqrt{5}
-1)/2 = 0.6180$ before declining to zero.

Equation~(\ref{expecloss}) shows that the loss ${\cal L}_n = (\operatorname{Var}
D_n)/n$. Expressions for $\operatorname{Var} D_n$ from the steady-state distribution
are given by \citet{markbill2010}, from which expressions for ${\cal
L}_n$ follow.

These values of ${\cal L}_n$ again depend on whether $n$ is even or odd:
%
\begin{eqnarray}
{\cal L}_{2m} &= &\frac{4r(r^2+1)}{n(r^2 - 1)^2}, \label{evenloss2}
\\
{\cal L}_{2m-1} &=& \biggl\{\frac{8r^2}{(r^2 - 1)^2} +1 \biggr\}\Big/n.
\label{oddloss}
\end{eqnarray}
Re-expressions as functions of $p$ do not provide any insight.

The left-hand panel of Figure~\ref{coinf1} shows the average values of
loss, $\bar{L}_n$, for $p = 2/3$. Unlike the plot of bias in the
right-hand panel, this plot shows only a small effect of the parity of
$n$. As is to be expected, the loss is larger for odd $n$. Since the
rule becomes random allocations as $p$ approaches 0.5 and sequential
balancing, Rule~D, as $p$ approaches one, it is to be expected that the
ratio of loss for odd to even $n$ increases with $p$: for $p = 0.55$
the ratio is 1.0002; for Efron's coin with $p = 2/3$ it has only risen
slightly to 1.0250, becoming 1.1333 when $p = 0.75$.

Perhaps of greater importance is the approach of the loss to the value
in \eqref{expecloss} when the steady-state variance is used. Table~2 of
\citet{markbill2010} gives exact values of the variance of $D_n$ for
$n$ both even and odd for a range of values of $p$. Convergence to the
asymptotic value is faster for larger values of $p$. For small $n$ and
$p$ near 0.5 the exact variance is appreciably smaller than that at the
steady-state.

\begin{figure}

\includegraphics{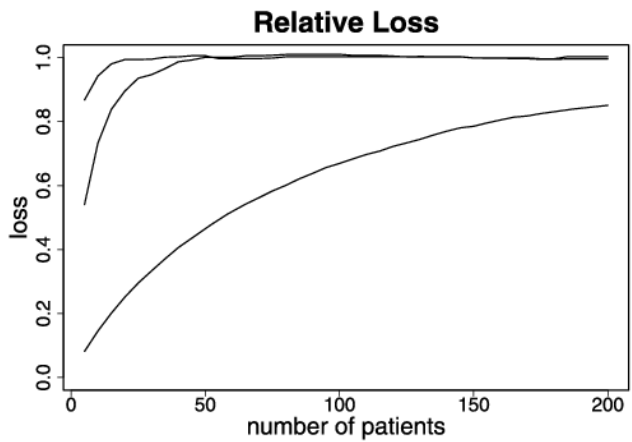}

\caption{Rule~E($2/3$). Convergence of ${\cal L}_n$ to the steady-state
value given by (\protect\ref{evenloss}) or (\protect\ref{expecloss}): ratio $n\bar{L}_n/D_{\infty}$. Reading down,
$p = 3/4$, $2/3$ and 0.55. 100,000
simulations.}\label{coinf13b}\vspace*{6pt}
\end{figure}

To illustrate the effect on loss, let the steady-state variance be
denoted $\operatorname{Var} D_{\infty}$. Then the loss should decrease as $1/n$ and
the ratio $n\bar{L}_n/D_{\infty}$ should tend toward one. Figure~\ref{coinf13b}
illustrates this for three values of~$p$. For $p = 0.55$ the
ratio has only reached 0.85 when $ n = 200$. However, for $p = 2/3$,
values around one are reached around $n = 50$. For $p = 3/4$ the values
are above 0.9 for $n \ge8$. The steady-state values of loss given in
\eqref{evenloss} and \eqref{oddloss} provide useful guidance for all
coins except those with very low values of $p$, which are unlikely to
be used in practice, or with small values of $n$.

These results about bias and loss indicate that the properties of
Efron's biased-coin are well understood, both asymptotically and for
smaller samples. In theory, exact results for ${\cal L}_n$ can be
obtained from the analytical expressions for the distribution of $D_n$
given in \citet{markbill2010}. However, the authors warn in their
Section~3 that care is needed in the numerical calculation of the
summations they present, since these involve factorials of large
numbers and powers of numbers less than one. As here, simulation may
sometimes be an easier way to obtain an idea of the properties of a
rule for a variety of parameter values.

\begin{figure*}

\includegraphics{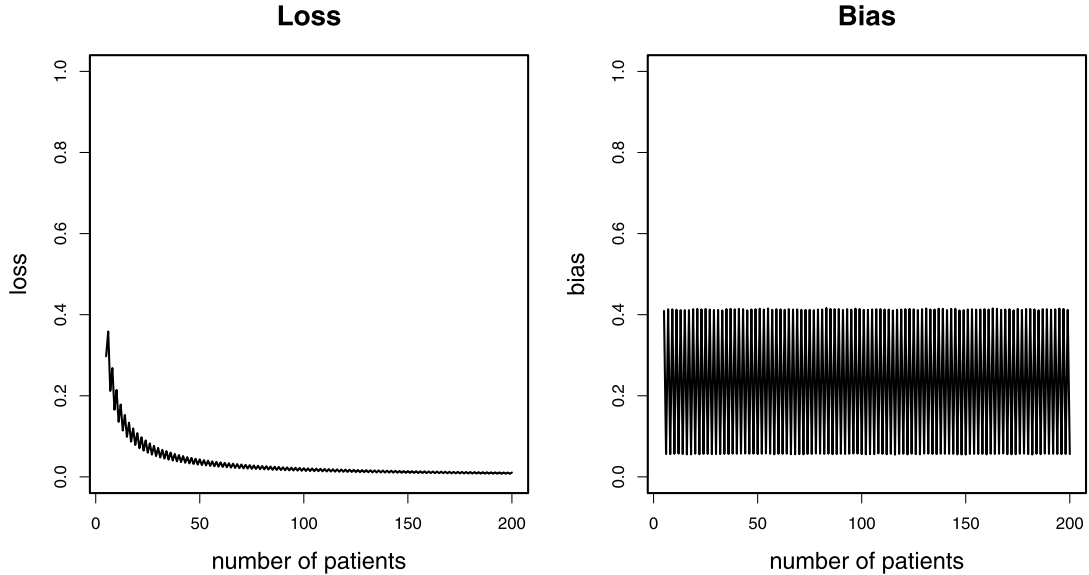}

\caption{Adjustable biased-coin with $a=3$---Rule J(3). Left-hand panel,
average loss $\bar{L}_n$, right-hand panel, average bias $\bar{B}_n$.
100,000~simulations.} \label{coinf3new}
\end{figure*}

\subsection{The Adjustable Biased-Coin: Rule~J}
\label{baldananusec}

Figure~\ref{coinf3new} shows the average values of loss $\bar{L}_n$ and
bias $\bar{B}_n$ for Rule~J with parameter $a = 3$. These plots are
similar in structure to those for Rule~E in Figure~\ref{coinf1},
although the values of loss are lower for Rule~J and show a greater
effect of the parity of $n$. The surprise, however, is the more extreme
values of bias, which are not to be expected from the presentation of
\citet{baldis2004}.

There are some theoretical results for the balance, bias and power of
Rule~J. \citet{baldis2004} provide asymptotic comparisons of
selection bias with Rule~E and show that their rule has smaller bias
than E($2/3$) for any value of $a$. The power comparisons of \citet
{baldi2008} are for a more general family of rules than J. One result
(Corollary 3) is that for any sample size a rule with $F(-1) \ge p$ is
uniformly more powerful that $E(p)$. The rule studied here, \eqref
{baldirule}, does not meet this condition. The rate of decrease of the
loss associated with Rule~J is shown by \citet{baldimz2011} to
be~$1/n$.

The adjustable biased-coin \eqref{baldirule} combines avoidance of
excessive imbalance with greater randomness than Efron's rule in the
centre of the distribution of $D_n$. For $D_n = -1, 0$ or 1 the
probability of moving to $D_{n+1}= D_n +1$ or $D_n - 1$ is 0.5; values
of $D_n = \pm2$ are therefore frequent. However, except for small $a$,
absolute values greater than or equal to four rarely occur---see the
numerical calculations in Tables~1--3 of \citet{baldis2004}. We
therefore use as an approximation a truncated Markov chain for $D_n$. For example, for $a = 2$, the equilibrium
probability that $|D_n| \ge4$ is less than 0.013 for $n$ even and less
than 0.001 for $n$ odd. To obtain a tractable approximation to the
equilibrium probabilities of this rule, and so to calculate approximate
values of the bias and loss, we examine the approximation of the rule
by the truncated Markov chain on the values $-3$ to 3. This
truncated rule is a hybrid between those of \citet{soareswu83} and of
\citet{ypchen99} introduced in Section~\ref{imtolsec}.

\begin{table}
\tabcolsep=0pt
\caption{Stationary distribution of the approximation to Rule~J; $p =
2^a/(1+2^a)$}\label{cointab1}
\begin{tabular*}{\columnwidth}{@{\extracolsep{\fill}}lccccccc@{}}
\hline
$\bolds{D_n}$ & $\bolds{-3}$ & $\bolds{-2}$ & $\bolds{-1}$ &
$\bolds{0}$ & $\bolds{1}$ & $\bolds{2}$ & $\bolds{3}$ \\
\hline
$n$ odd & $\frac{1-p}{2(1+p)}$ & 0 & $\frac{p}{1+p}$ & 0 & $\frac
{p}{1+p}$ & 0 & $\frac{1-p}{2(1+p)}$ \\
$n$ even & 0 & $\frac{1}{2(1+p)}$ & 0 & $\frac{p}{1+p}$ & 0 & $\frac
{1}{2(1+p)}$ & 0 \\
\hline
\end{tabular*}
\end{table}
The stationary distribution of $D_n$ for this approximation is
displayed in Table~\ref{cointab1} where $p = 2^a/(1+2^a)$. Given the
stationary distribution, it is straightforward to calculate the loss
and bias.
The loss for a difference of $D_n$ is $D^2_n/n$. The biases for the
central values of $D_n$, that is, 0 and $\pm1$, are zero, since the
allocation is at random. For $|D_n| = 2$ there is a probability of $p$
of allocating the underrepresented treatment and the conditional bias
is $2p-1$. For $|D_n| = 3$ the underrepresented treatment is always
allocated and the bias is one. Taking expectations over the stationary
distributions for odd and even $n$ leads to
%
\begin{eqnarray}
\label{approxbl}&& n \mbox{ even:}\quad {\cal L}_n = \frac{9-7p}{n(1+p)},\quad  {\cal B}_n = \frac{2p-1}{1+p},
\nonumber
\\[-8pt]
\\[-8pt]
\nonumber
&&
n \mbox{ odd:}\quad {\cal L}_n = \frac{4}{n(1+p)}, \quad {\cal
B}_n = \frac{1-p}{1+p}.
\end{eqnarray}
As for Rule~E, for this approximation the expected loss decreases with
$n$, but the bias is independent of $n$, in line with the plots of
Figure~\ref{coinf3new}.

\begin{table}
\caption{Rule~J. Approximate values of ${\cal L}_n$ and ${\cal B}_n$
($n = 199$ and 200) from the approximation \protect\eqref{approxbl}. Bracketed
values below are averages $\bar{B}_n$ and $\bar{L}_n$ from 100,000 simulations}\label{tab2}
\begin{tabular*}{\columnwidth}{@{\extracolsep{\fill}}lcccc@{}}
\hline
$\bolds{a}$ & $\bolds{L_{199}}$ & $\bolds{L_{200}}$ & $\bolds{B_{199}}$ & $\bolds{B_{200}}$ \\
\hline
1 & 0.0131 & 0.0120 & 0.2000 & 0.2000 \\
& (0.0172) & (0.0177) & (0.2369) & (0.1382) \\
2 & 0.0095 & 0.0111 & 0.3333 & 0.1006 \\
& (0.0100) & (0.0120) & (0.3408) & (0.1006) \\
3 & 0.0074 & 0.0106 & 0.4118 & 0.0588 \\
& (0.0075) & (0.0107) & (0.4152) & (0.0579) \\
4 & 0.0062 & 0.0103 & 0.4545 & 0.0303 \\
& (0.0062) & (0.0103) & (0.4545) & (0.0303) \\
\hline
\end{tabular*}    \vspace*{3pt}
\end{table}

\begin{figure*}[b]

\includegraphics{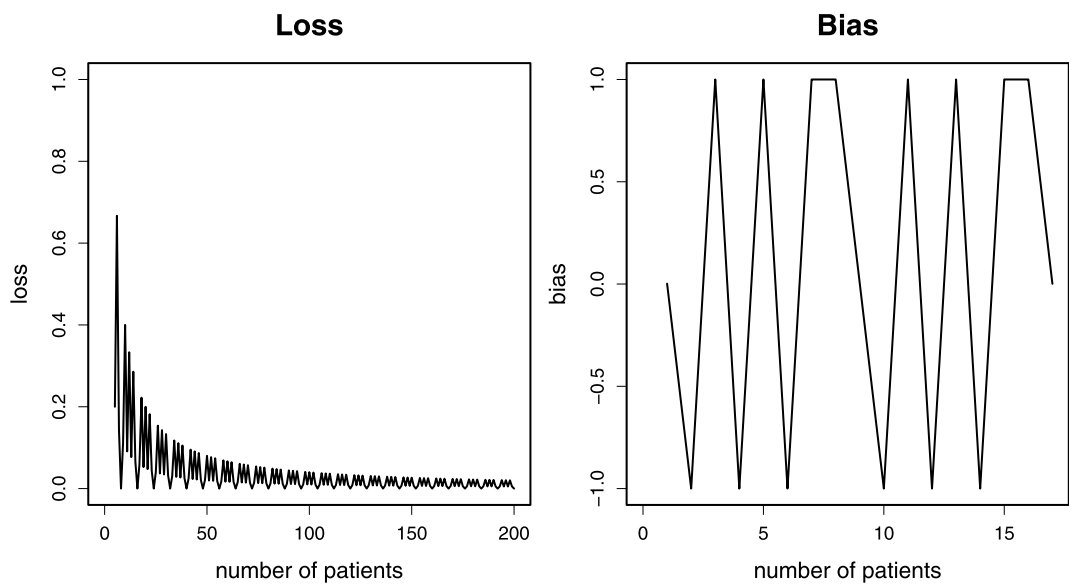}

\caption{Permuted block design, block size eight---{Rule P}. Left-hand
panel, loss ${\cal L}_n$, right-hand panel, bias ${\cal B}_n$; the value
of $-1$ comes from incorrect guessing.}\label{cap2f8}
\end{figure*}

Table~\ref{tab2} gives a comparison of the values of ${\cal L}_n$ and
${\cal B}_n$ from the approximation \eqref{approxbl} with, in brackets,
the average values from 100,000 simulations for $n = 199$ and 200. The
table shows how good the approximation is, both for bias and loss, when
$a$ is as small as 2, that is, $p = 4/5$. For this and higher values of
$a$, the value of $p$ is sufficiently large that the distribution of
$D_n$ is indeed concentrated in the range $-3$ to 3. Better
approximations for both small $a$ and small $n$ can be found by putting
the reflecting barriers of the Markov process further out
than~$\pm3$.\looseness=1

There are three substantive points in these results. The first is the
extremely small values of loss, even when $a = 1$. For these values of
$n$ an arbitrarily stopped trial will be very close to balance. The
second is that, although the values of loss depend on the parity of
$n$, the values are so small that the inferential effect is negligible.
The third is the extremely high value of bias when $n$ is even. As $p
\rightarrow1$ the approximations in \eqref{approxbl} show that the
bias tends to 0.5 for $n$ even, whilst going to zero for $n$ odd. This
behaviour raises the question of how to compare rules which have such
different properties for odd and even $n$.

\subsection{Rule~P: Permuted Block Design}

As an example of a permuted block design let $n = 8$. A typical design
allocates treatments in the order AABABABB. The underrepresented
treatment is guessed with random guessing for the first
allocation. Then ${\cal B}_1 = 0$.
Thereafter, ${\cal B}_n$ will be one when the underrepresented
treatment is allocated and $-1$ when the overrepresented treatment
is allocated. If the length of the block is known, the last guess
will always be correct, as balance is attained. For example,
guessing the underrepresented treatment in AABABABB gives ${\cal
B}_2 = -1$ and ${\cal B}_8 = 1$.

\begin{figure*}

\includegraphics{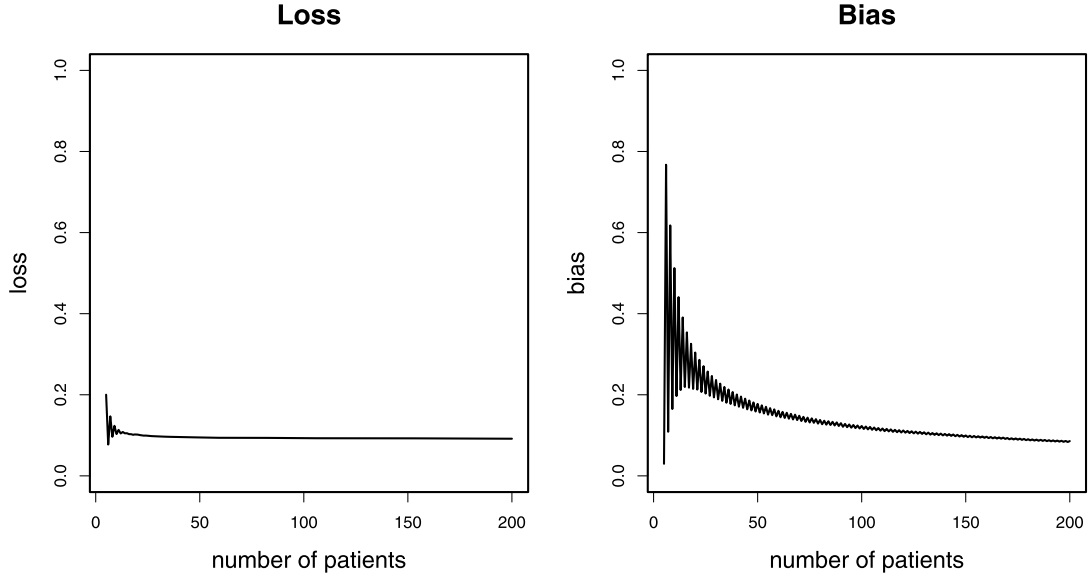}

\caption{Smith's rule with $\rho= 5$---Rule S(5). Left-hand panel,
average loss $\bar{L}_n$, right-hand panel, average bias $\bar{B}_n$.
100,000 simulations.}\label{coinf5}
\end{figure*}

Figure~\ref{cap2f8} shows loss and bias for the rule with block size 8.
Loss quickly decreases with $n$; since there is balance when each block
is completed, ${\cal L}_{8m} = 0$. Because the allocation is
deterministic, the fine detail of the plot shows repetition of the same
eight-allocation pattern of loss, decreasing as $1/n$. The right-hand
panel shows bias up to $n = 16$, that is, two cycles of guessing in
ignorance of the structure.

Figure~1 of \citet{efr71} compares a measure of selection bias for
several biased-coin Rules E with permuted blocks of size $2m$. For $ m
= 9$ the bias is similar to that of E($2/3$). As $m$ increases, the rule
becomes more like random allocation, that is, Rule~E with $p
\rightarrow0.5$. Figure~6.3 of \citet{rosl2002} compares bias over
$n$ for values of $m$ from 1 (deterministic allocation) to 5. These
comparisons are over all possible permutations of the $2m$ allocations,
rather than for a specific permutation, like that of Figure~\ref{cap2f8} that would be used in a particular trial. \citet{rabpn2003}
advocate restricted randomization, in which permutations with an
``obvious'' pattern are not considered. Those that become too unbalanced
could also be excluded.

The ability to guess correctly depends on what is known about the
structure of the design. If it were known that this structure of
eight treatments were to be repeated, then ${\cal B}_n$ would be one
for all $n > 8$. Randomly relabeling treatments A and B, using
several permutations or changing the block size are all ways in
which the value of ${\cal B}_n$ could be kept small, although at some
administrative cost.

\subsection{Smith's Rule: Rule~S}
\label{smithananumsec}

Figure~\ref{coinf5} shows the values of average loss and average bias
for 100,000 simulations of Smith's rule with $\rho= 5$. Apart from the
few initial values of $n$, the values of loss in the left-hand panel
are virtually constant, whereas the average bias decreases with $n$. As
it does so, the effect of the parity of $n$ disappears. This is very
different behaviour from that for the two-biased coin Rules J and E in
which bias is constant with $n$, although depending on parity, whilst
loss decreases as $1/n$.

The properties of the biased-coin designs in the earlier sections were
found from the Markov chain formed by the values of $D_n$. This
structure is not available for Rule~S and asymptotic arguments are used
instead. From equation (4.1) of \citet{rls84b},
%
\begin{equation}
{\cal L}_n = 1/(1 + 2\rho). \label{smithloss}
\end{equation}
The asymptotic distribution of ${L}_n$ follows from \citet{rls84a} who
shows the convergence in distribution of $n^{-1/2} D_n \rightarrow
{\cal N}\{0,1/(1+2\rho)\}$. From \eqref{expecloss} the asymptotic
distribution of loss is therefore
\[
L_n \sim X_1^2/(1 + 2\rho),
\]
where $X_1^2 \sim\chi^2_1$.

The asymptotic distribution of $D_n$ also provides the asymptotic
distribution of $N_i$, the number of patients receiving treatment $i$.
Since $\operatorname{Var} (N_i) = \operatorname{Var} (D_n)/4 $, asymptotically,
%
\begin{equation}
n^{-1/2} N_i \sim{\cal N}\bigl[1/2, 1/\bigl\{4(1+2\rho)
\bigr\}\bigr]. \label{dibnni}
\end{equation}
For random allocation ($\rho= 0$) the variance is $1/4$.

In his (4.3) Smith further uses the asymptotic normality of $D_n$ to
show that, as $n \rightarrow\infty$,
%
\begin{equation}
{\cal B}_n \cong\rho\sqrt{2/\bigl\{n \pi(1 + 2\rho)\bigr\}}.
\label{smithbias}
\end{equation}

\begin{figure*}

\includegraphics{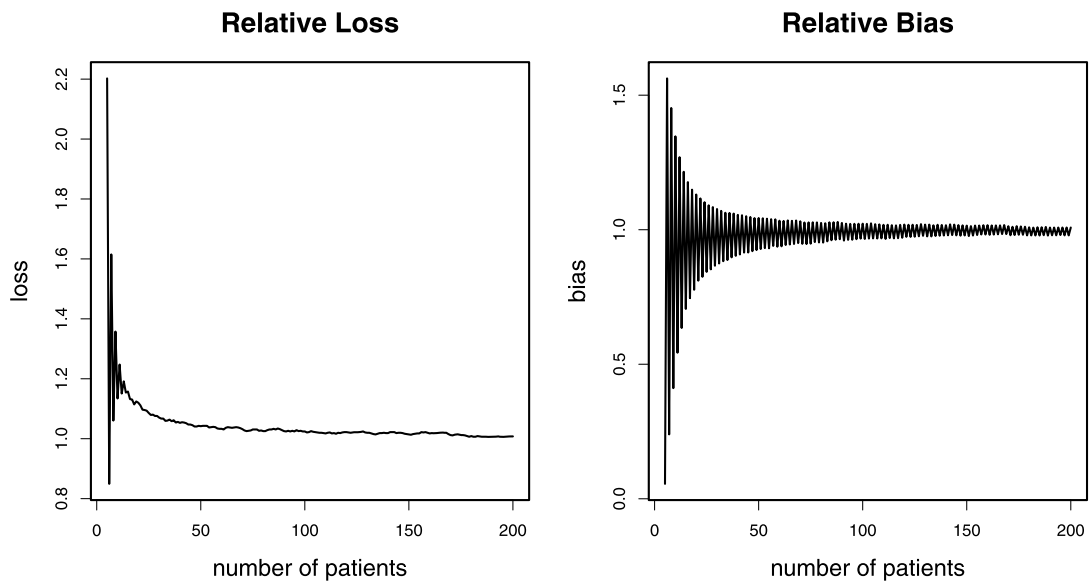}

\caption{Rule~S(5). Approach of loss and bias to asymptotic values.
Left-hand panel, ratio of $\bar{L}_n$ to ${\cal L}_n$ (\protect\ref{smithloss}). Right-hand
panel, ratio of $\bar{B}_n$ to ${\cal B}_n$ (\protect\ref{smithbias}). 100,000 simulations.}
\label{coinf14}
\end{figure*}

Figure~\ref{coinf14} explores the relationship between the average
values of loss and bias plotted in Figure~\ref{coinf5} and the
asymptotic values given above. The left-hand panel shows the ratio of
the two estimates of expected loss, $\bar{L}_n$ and ${\cal L}_n$.
Initially there is a slight effect of the parity of $n$ on the ratio,
but, from $n = 15$, the ratio decreases from less than 1.2 toward one
as $n$ increases. The plot of the ratio for bias, $\bar{B}_n/{\cal
B}_n$ in the right-hand panel, shows the much stronger effect of odd
and even $n$ which was also apparent in Figure~\ref{coinf5}, but is
ignored in the asymptotic expression \eqref{smithbias}. The ratio is centered on one with the effect of
parity steadily decreasing.

The indication of Figure~\ref{coinf5} is that the asymptotic results
for Rule~S provide a good guide to the behaviour of this rule, even for
small values of $n$.

\subsection{Bayes: Rule~B}
\label{bayesananumsec}

Unlike the other rules of this section, there are no theoretical
results for the loss and bias of Rule~B, except that it moves from
deterministic allocation, Rule~D, to random allocation as $n$
increases. The rate of transition from one form of allocation to the
other depends on the value of the parameter $\gamma$.

\begin{figure*}

\includegraphics{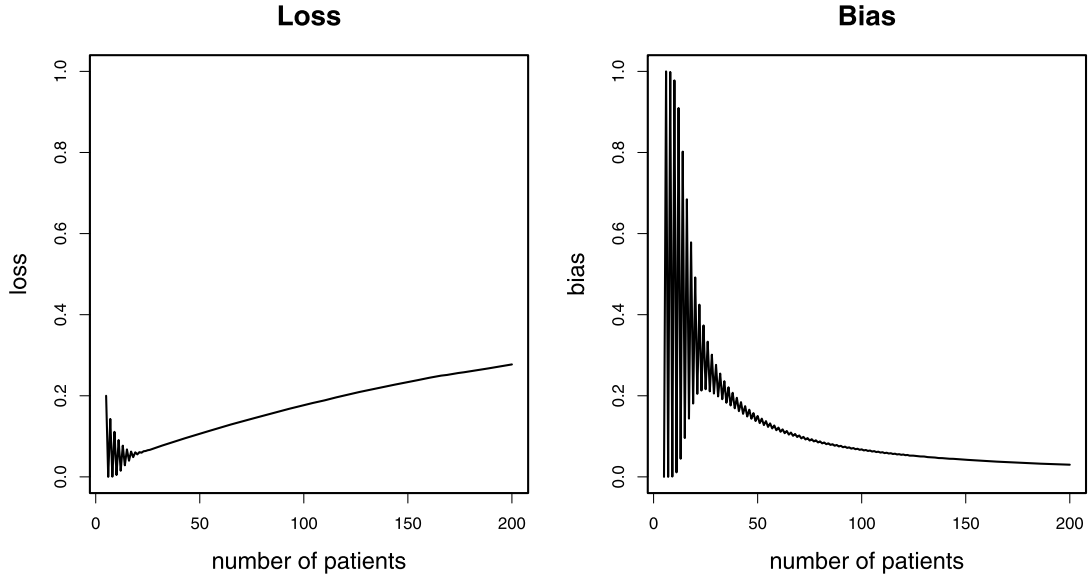}

\caption{Bayes rule with $\gamma= 0.01$---Rule B(0.01). Left-hand
panel, average loss $\bar{L}_n$, right-hand panel, average bias $\bar
{B}_n$. 100,000 simulations.}\label{coinf4}\vspace*{-3pt}
\end{figure*}

The final plot of bias and loss for a single rule without covariates is
in Figure~\ref{coinf4} for the Bayes rule with $\gamma= 0.01$. These
plots are unlike any we have so far seen. The left-hand panel shows
that the loss starts close to zero and then gradually increases with
$n$. Initially, the balancing effect of the design is such that the
loss depends on the parity of $n$. The bias, in the right-hand panel,
starts by alternating between zero and one, as it does for
deterministic allocation. As $n$ increases the bias decreases, as the
rule becomes increasingly like random allocation.\looseness=1

\section{Adjacent Averages}
\label{adjavsec}

A striking feature of Figures~\ref{coinf1} and \ref{coinf3new} is the
strong dependence of the bias on whether $n$ is even or odd, although
this is not a feature of all rules. In the next section we compare
several of the rules of Section~\ref{ananumsec} for various parameter
values. Table~\ref{tab3} extends the simulation results of Table~\ref{tab2} to these nine rules.

\begin{table*}[b]
\caption{Average values of loss, $\bar{L}_n$, and bias, $\bar{B}_n$ ($n
= 199$ and 200), and adjacent
averages of these values, $\tilde{L}_{200}$ and $\tilde{B}_{200}$, for
nine rules arranged according to adjacent average bias. 100,000
simulations}\label{tab3}
\begin{tabular*}{\textwidth}{@{\extracolsep{\fill}}lcccccc@{}}
\hline
\textbf{Rule} & $\bolds{\bar{L}_{199}}$ & $\bolds{\bar{L}_{200}}$ &
$\bolds{\bar{B}_{199}}$ & $\bolds{\bar
{B}_{200}}$ & $\bolds{\tilde{L}_{200}}$ & $\bolds{\tilde{B}_{200}}$ \\
\hline
D & 0.0050 & 0.0000 & 0.0022 & 1.0000 & 0.0025 & 0.5011 \\
E($2/3$) & 0.0228 & 0.0221 & 0.1707 & 0.3371 & 0.0224 & 0.2549 \\
J(3) & 0.0075 & 0.0107 & 0.4152 & 0.0579 & 0.0091 & 0.2366 \\
E(0.55) & 0.2139 & 0.2127 & 0.0848 & 0.1041 & 0.2133 & 0.0944 \\
S(5) & 0.0916 & 0.0916 & 0.0861 & 0.0874 & 0.0916 & 0.0868 \\
S(2) & 0.2001 & 0.2002 & 0.0491 & 0.0518 & 0.2002 & 0.0505 \\
B(0.01) & 0.2764 & 0.2773 & 0.0279 & 0.0313 & 0.2769 & 0.0296 \\
B(0.1) & 0.6972 & 0.6982 & 0.0050 & 0.0032 & 0.6917 & 0.0041 \\
R & 1.0010 & 1.0007 & 0.0022 & 0.0025 & 1.0008 & 0.0024 \\
\hline
\end{tabular*}
\end{table*}

In the table the rules are arranged in order of decreasing bias.
The first four columns are the average losses and biases from 100,000
simulations when $n = 199$ and 200. As the figures for individual rules
have shown, the values of bias are more sensitive to the parity of $n$
than are the values of loss. The strongest difference in bias between
$n$ odd and $n$ even is for deterministic allocation, Rule~D, when the
theoretical value of the bias is one or zero depending on whether $n$
is even or odd.

We have already seen in Section~\ref{efrsec} that the expected values
of bias for Efron's rule, E($2/3$), are $1/6$ and $1/3$. For the adjustable
biased-coin J(3), the results of Table~\ref{tab2} are that the two
values of bias are 0.4118 and 0.0588. As $p$ decreases toward 0.5 in
Efron's rule, the allocation becomes more random and the difference in
properties for odd and even $n$ decreases. When $p = 0.55$ the expected
value of bias for $n =199 $ and 200 are 0.0818 and 0.1. The remaining
rules in the table all have values of bias less than 0.1.

Trials are equally likely to stop with $n$ odd or even. Therefore, in
order to compare rules, we use adjacent averages and write
%
\begin{equation}
\tilde{L}_n = (\bar{L}_{n-1} + \bar{L}_{n})/2,
\label{adjloss}
\end{equation}
with a similar definition for $\tilde{B}_n.$ These averages remove the
effect of oscillation between the two values, particularly of bias, and
allow an insightful comparison of rules.

For the rules with continuous covariates compared in Section~\ref{normalsec} the maximum value of ${\cal B}_n$ is one. But, without
covariates, the most extreme adjacent values of bias are zero and one,
so that, with adjacent averaging, the maximum value of E$ \tilde{B}_n$
is 0.5. The rules in Table~\ref{tab3} are arranged in decreasing order
of $\tilde{B}_{200}$. The largest theoretical value is 0.5 for Rule D.
The only other rules with appreciable averages for adjacent bias
are\vadjust{\goodbreak}
E($2/3$) and J(3) with values around 0.25; for all other rules the values
are less than 0.1.

It is a general principle in the comparison of these rules that
acceptable rules with high bias have low loss and conversely. \citet
{aca2002} provides examples for allocation rules with covariates. And,
indeed, in Table~\ref{tab3} the values of $\tilde{L}_{200}$ increase
from a theoretical value of 0 for Rule~D to 1 for Rule~R. The only
exceptions are the two Efron rules; E($2/3$) has higher loss and higher
bias, at $n = 200$, than J(3). Similarly, E(0.55) has higher loss and
bias than either of the Rules S. These results, however, only provide a
snapshot of the behaviour of the rules at $n = 200$. We now look at
plots of the adjacent averages of bias and loss over a range of values
of $n$.

\begin{figure*}[t]

\includegraphics{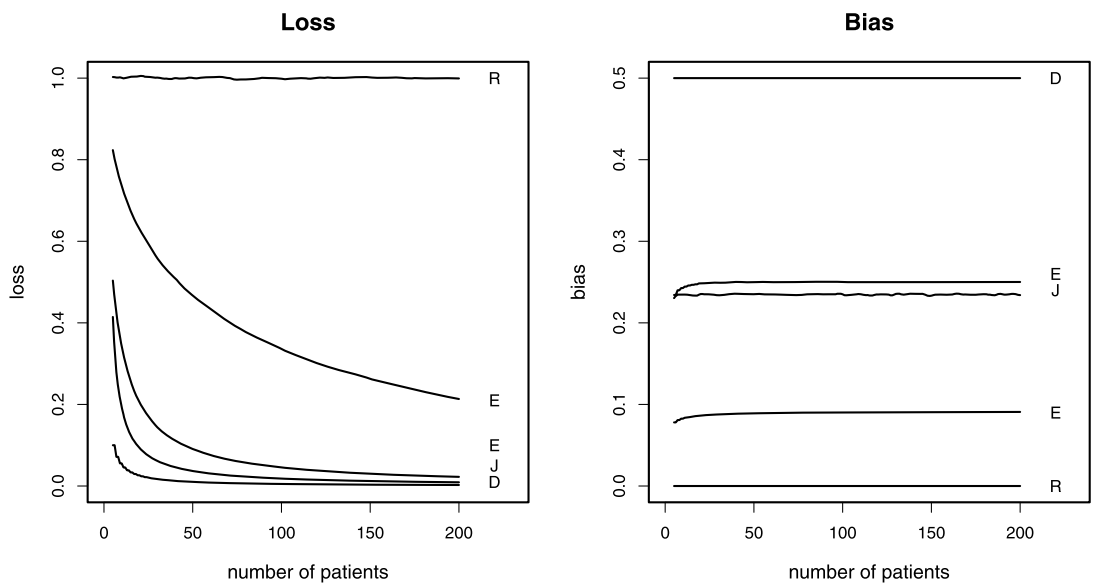}

\caption{Simulated values of adjacent averages of loss and smoothed
bias for five rules. Left-hand panel, $\tilde{L}_n$. Reading down: R,
random allocation; E(0.55), Efron's rule with $p = 0.55$; E($2/3$);
adjustable biased-coin J(3) and D, deterministic allocation. Right-hand
panel, $\tilde{B}_n$. Same notation, reading up. 100,000 simulations.}
\label{coinf10}\vspace*{-3pt}
\end{figure*}

Figure~\ref{coinf10} shows plots of adjacent averages of bias and loss
for the rules in which loss decreases as $1/n$ and the bias is
constant, that is, Rules J and E. Also included are random and
deterministic allocation, which are the special cases of Rule~E as $p
\rightarrow0.5$ and one. The values of $\tilde{L}_n$ in the left-hand
panel form a series of values of loss decreasing steadily with $n$.
Likewise, the right-hand panel shows a series of virtually constant
values for $\tilde{L}_n$. The only surprise is that Rule~J(3) has lower
loss and lower bias than E($2/3$). The argument of the next section is
that Rule~E($2/3$) is therefore inadmissible. Otherwise, the order of the
rules is reversed between the panels for loss and bias.

\begin{figure*}

\includegraphics{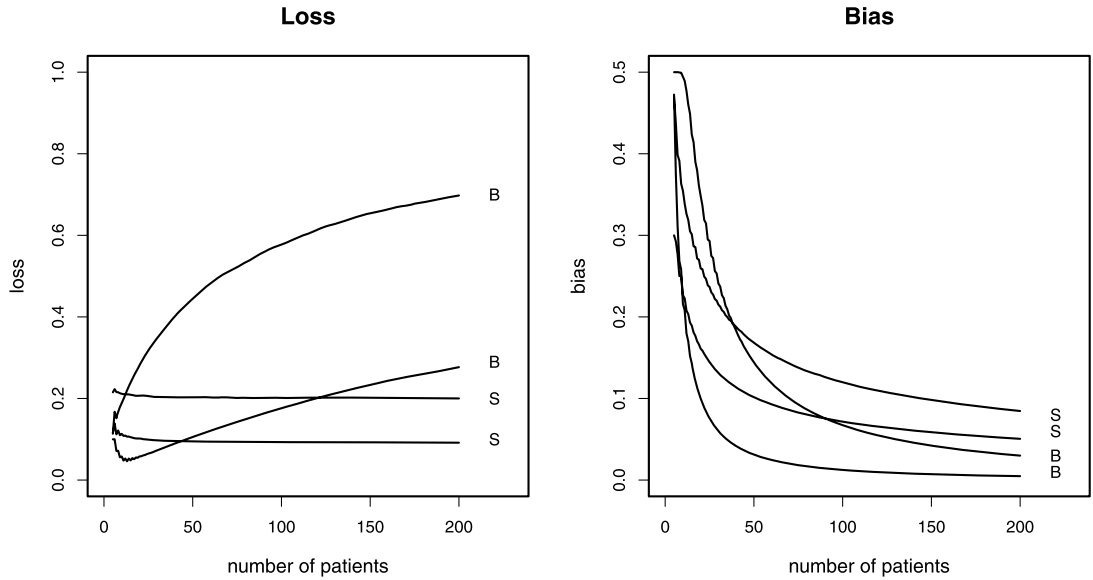}

\caption{Simulated values of adjacent averages of loss and smoothed
bias for four rules. Left-hand panel, $\tilde{L}_n$. Reading down: B,
Bayes rule with $\gamma= 0.1$; B, Bayes rule with $\gamma= 0.01$; S,
Smith's rule with $\rho= 2$ and Smith's rule with $\rho= 5$.
Right-hand panel, $\tilde{B}_n$. Same notation, reading up. 100,000
simulations.}\label{coinf11}
\end{figure*}

The plots of $\tilde{L}_n$ for Rule~S in the left-hand panel of
Figure~\ref{coinf11} are virtually constant, whereas those for Rule~B
increase with $n$. As $n$ increases, Rule~B$(0.1)$ becomes increasingly
like random allocation at a faster rate than does the rule with $\gamma
= 0.01$. The plots for $\tilde{B}_n$ in the right-hand panel are in the
reverse order by the time $n$ is around 100; as the loss for Rule~B
increases, the bias decreases faster than $1/n$ as allocation becomes
increasingly random.

\section{Admissibility}
\label{admisssec}

A good rule should have low loss and low bias for all~$n$. In order to
compare loss and bias, \citet{aca2002} suggested plotting loss against
bias as a function of $n$, thus combining in one plot the two panels of
plots like Figure~\ref{coinf11}. The shape that this curve takes will
depend upon the individual rule; from the curves in Figures~\ref{coinf10} and \ref{coinf11} it is clear that there will be three
general forms of trajectory. A good rule will be in the lower left-hand
corner of the plot. But usually, for any particular $n$, a rule with
lower loss than another will have higher bias. A rule for which both
values are higher is said to be inadmissible.

\begin{figure*}[b]

\includegraphics{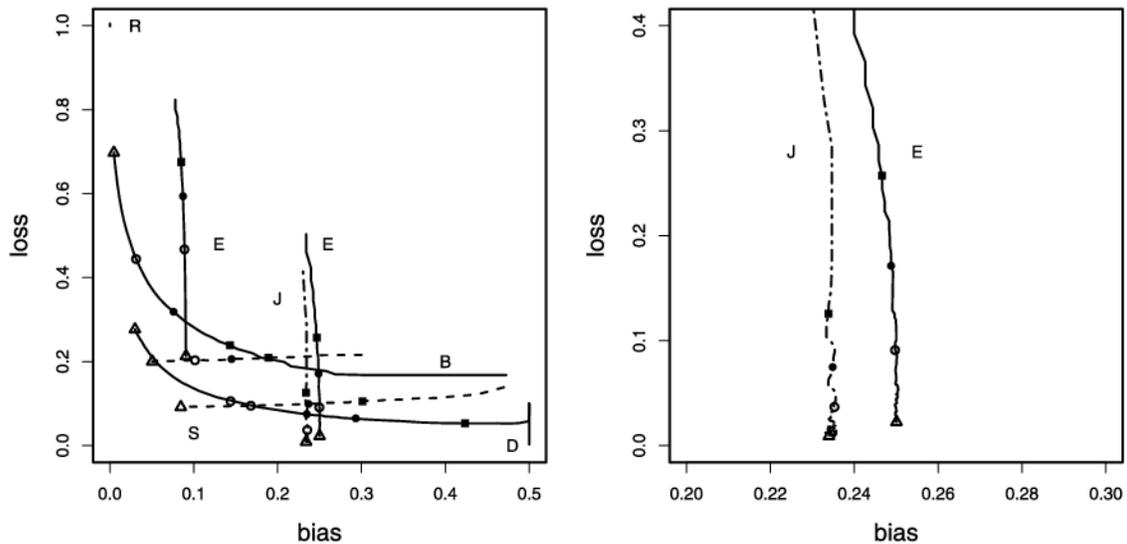}

\caption{Admissibility. Simulated values of adjacent averages of loss
and smoothed bias for nine rules. Left-hand panel, vertical continuous
lines from the left---E(0.55) and E($2/3$); vertical dotted and dashed
line---J(3); horizontal dashed lines reading down---S(2) and S(5);
continuous lines, reading down---B(0.1) and B(0.01). Symbols:
$\blacksquare\ n = 15; \bullet\ n = 25$;  $\circ\ n = 50$; $\Delta\ n =
200.$ Right-hand panel, J(3) and E($2/3$). 100,000 simulations.}\label{coinf12}
\end{figure*}

The left-hand panel of Figure~\ref{coinf12} is the admissibility plot
for the nine rules considered in this paper. Rules~R and D plot
virtually as points at $(0,1)$ and $(0.5,0)$, respectively. Both are
admissible since 0 is the minimum possible value of bias or loss. The
other rules in Figure~\ref{coinf10}, which have constant bias, plot as
vertical lines. The symbols on the lines correspond to values of $n$
with $\Delta$ denoting $n = 200$. For E(0.55) the bias is around 0.09
while the loss decreases steadily with $n$. For Rules~J(3) and E($2/3$)
the bias is higher but the loss is lower for specific $n$. Rule~E(0.55)
is therefore admissible when compared with these two rules. However,
the comparison of Rules~J(3) and E($2/3$) in the right-hand panel of the
figure shows that E($2/3$) is not admissible. For any $n$ over the range
10 to 200 it is possible to achieve lower loss and bias by using Rule~J(3).

The Rules S in Figure~\ref{coinf11} have virtually constant loss and so
plot as almost horizontal lines; S(5) plots below S(2) as it has lower
loss, but it lies to the right of the curve for S(2) for any particular
$n$. As $n$ increases the bias decreases and the curves tend toward
the left-hand axis. The curves for Rule~B form a third family. These
start close to Rule~D, but the bias decreases as the loss increases.
For $n = 10$ they have the smallest loss for all rules except D and,
for $n = 200$ the lowest bias for all rules except R. For smaller
values of $\gamma$ than these, the emphasis in Rule~B is increasingly
on balancing the design and a series of curves is obtained which lie
below those plotted. However, for any $n$, the points are to the right
of those for larger $\gamma$ so that all these rules are admissible.

The right-hand panel of Figure~\ref{coinf12} shows that E($2/3$) is not
admissible compared to J(3) over all values of $n$ considered. The
comparison of results in Table~\ref{tab3} indicated that, for $n =
200$, Rule~E(0.55) had higher bias and loss than Rules~S. But the
left-hand panel of the figure shows that, for $n$ less than 50, the
bias of S(2) is greater than that of E(0.55). Further, the indication
is that, for $n$ slightly greater than 200, E(0.55) has lower loss than
S(5). In addition, for almost all $n$ in the figure, S(2) has higher
bias than E(0.55). Neither rule dominates the other over all considered
values of $n$.

Superficially, Figure~\ref{coinf12} appears similar to several of the
figures in \citet{zhao2011} where a measure of bias is plotted
against maximum absolute imbalance, that is, the maximum of $|D_n|$ in
each simulation. However, the plots are for fixed  even $n$ for a
series of parameter values. Information on the dependence on $n$ is not
clear. As Figure~\ref{coinf3new} for Rule~J shows, very different
conclusions can be reached about the properties of a rule if only odd
or even values of $n$ are considered. Section~\ref{asesssec} emphasizes
the statistical basis of the criteria, selection bias and loss, used
here. Use of the maximum of $|D_n|$ favours rules in which the values
of $D_n$ are bounded and, indeed, the rule of \citet{soareswu83},
which is random over a restricted range, performs best in the
comparisons of \citet{zhao2011}; the rule of \citet{baldis2004} is
not included in their comparisons.

\section{Designs with Covariates}
\label{covarsec}
When measurements of some covariates are available before treatment
allocation, the randomization of patients should allow for the
covariates. \citet{rossverd2008} present a survey of the approaches of
statisticians and clinical trialists to the handling of covariates in
the design of clinical trials. Their numerical examples are for
binomial responses, which are naturally heteroscedastic. They stress,
and illustrate by example, that under such conditions, balance over
covariates does not lead to the most efficient designs. In contrast,
the next sections of this paper continue the study of rules for normal
responses with constant variance. Under these conditions balance of
covariates over treatments reduces dependence on the correctness of the
assumed form of the relationship between response and covariates. Even
if the covariates are to be adjusted for in the analysis, balance
ensures estimates of effects with lower variance \citep{baldimz2011}.
Comments on designs for heteroscedastic models and generalized linear
models are given briefly in Section~\ref{concsec}.

Comparisons of the loss, bias and admissibility of several rules with
covariates and normally distributed errors are given by \citet
{aca2002} and in Chapter~6 of \citet{acaatanu2014a}. However, these
comparisons do not include Rule~J. Accordingly, the focus here is on
extensions of this rule to include covariates.

\textit{Rule~M: Minimization---Pocock and Simon.}
We start with two deterministic rules which do not model the dependence
of the response on the covariates. The minimization rule of \citet
{ps75} depends on calculating the total effect on all measures of
marginal imbalance when treatment $j^+$ is allocated. If there are $m$
covariates $x$, there will be $m$ measures to be summed. The individual
measures count the number of observations in each category of the
covariate. Continuous covariates therefore have to be categorized.

Let the total effect on imbalance be $C(j^+)$. The allocations are
ranked so that
\[
C\bigl([1]\bigr) \le C\bigl([2]\bigr).
\]
In this deterministic allocation treatment [1] is allocated, with
random allocation if both treatments have the
same value of $C(j^+)$.

The calculations are exemplified by \citet{sennval2010} and \citet
{aca2002} as well as by \citet{ps75}. In the simulations of
Section~\ref{normalsec} standard normal covariates are dichotomized at~0.

\textit{Rule~C: Balanced covariates.}
There are $m$ covariates, either discrete or discretized, with
covariate $i$ having $l_i$ levels. The total number of cells or strata
is $M = \prod_{i=1}^ml_i$. Suppose
that the covariate vector $x_n$ for patient $n$ falls in cell $\iota$.
The new allocation
depends solely on previous allocations in that cell. Balance is most
effectively forced
by using deterministic allocation independently within each
of the $M$ cells. If there are any ties, random allocation is used.

\textit{Randomized versions of Rules M and C}.
\textit{Rule~ME.} Randomization can be introduced into Rule M by
allocation of the treatments with probabilities given by the
biased-coin of Efron applied to the ordered values
of the $C(j^+)$;
\[
\pi_{\mathit{ME}}\bigl([1]|x_{n+1}\bigr) =2/3,
\]
again with random allocation if there is a tie.

\textit{Rule~CE}. Once the covariate cell $\iota$ has
been identified in Rule~C, the allocation within that cell is deterministic.
Rule~CE is a randomized version of Rule C, using Rule~E for allocation
within each cell.

\textit{Rule~CJ.} \citet{baldimz2011} suggest the rule in which the
adjustable biased-coin without covariates, Rule~J, is applied to the
numbers of times each
treatment has been allocated in cell $\iota$. They provide an
expression for loss and show that, for discrete covariates, $L_n
\rightarrow0$, a result which also applies to the nonrandomized Rule~C. They further discuss conditions under which marginal balance does
not guarantee global balance over all strata.

The other approach is to use measures from the optimum design of
experiments to determine the ``underrepresented'' treatment.
In an extension of the model of Section~\ref{asesssec} it is assumed
that the observations $y_i$ will be analysed using a regression model.
Now patient $i$ presents with a vector of covariates $x_i$. The
response is modelled using a vector of $q-1$ explanatory variables
$z_i$, to allow for any necessary interactions, quadratic terms and so
on of the $x_i$ which may be expected to be important. The parameter of
interest is still the treatment difference $ \mu_1 - \mu_2$, with a
vector $\psi$ of regression parameters not of importance, although
balance is wanted over these variables. Together with the mean response
$\beta_0$ there are then $q$ nuisance parameters. The model for all $n$
observations, in matrix form, is
%
\begin{equation}
\mathrm{E} Y = a \Delta+ 1 \beta_0 + Z\psi= a \Delta+ F \beta= G
\omega, \label{5aca6}
\end{equation}
where $\Delta= (\mu_1 - \mu_2)/2$ and $a$ is the $n \times1$ vector
of allocations with elements $+1$
and $-1$, depending on whether treatment 1 or treatment 2 is
allocated. The constant term and covariates are included in the
$n \times q$ matrix $F$. The value of $q$ is important in determining
the loss for some rules.

In sequential treatment allocation the covariates and allocations are
known for the first $n$ patients, giving a matrix $G_n$ of allocations
and explanatory variables in \eqref{5aca6}. Let patient $n+1$ have a
vector $z_{n+1}$ of explanatory variables. If treatment $j$ is
allocated, the vector of allocation and explanatory variables for the
$(n+1)$st patient is $g_{j,n+1},  j = 1,2$. Results in the sequential
construction of optimum experimental designs (\cite{aca82}, \cite{rls84b},
Section~10) show that the variance of the estimate $\hat
{\Delta}$ after $n+1$ observations is minimized by the choice of that
treatment for which the derivative function
%
\begin{eqnarray} \label{5defd}
d(j,n,z_{n+1}) &=& g^{\mathrm{T}}_{j,n+1}\bigl(G_n^{\rm T}G_n
\bigr)^{-1} g_{j,n+1}
\nonumber
\\[-8pt]
\\[-8pt]
\nonumber
&&{}- f^{ \mathrm{T}}_{j,n+1}
\bigl(F_n^{\mathrm{T}}F_n\bigr)^{-1}
f_{j,n+1}
\end{eqnarray}
is a maximum. See \citet{ADT2007}, Section~10.3, with $s = 1$.

The loss from randomization is assessed from $\operatorname{Var}(\hat{\Delta})$. Let
$b = F^\mathrm{T}a$, a
``balance'' vector which is identically zero when all
covariates are balanced across all treatments. Then
%
\begin{equation}
\operatorname{var} (\hat{\Delta}) = \frac{\sigma^2}{n - b^\mathrm{T}(F^{\mathrm{T}}F)^{-1}b} = \frac{\sigma^2}{n - {L}_n},
\label{aca8}
\end{equation}
giving an explicit expression for calculation of $L_n$. The loss is
minimized for the balanced design when the estimate of $\Delta$ is
independent of the estimates of the nuisance parameters. For a careful
discussion of the balance induced by allocation rules see \citet{baldimz2011}.

Asymptotic results on the distribution of $L_n$ are available for Rule~S. \citet{bur96} shows, following \citet{rls84b}, Section~10, that
%
\begin{equation}
L_n \sim X^2_q/(1 + 2\rho), \label{losschisq}
\end{equation}
where $X^2_q \sim\chi^2_q$. Thus, for random allocation $(\rho=
0),\break
{\cal L}_{\infty} = q$, the number of
nuisance parameters. For Atkinson's original proposal of D$_{{A}}$-optimality \mbox{$(\rho= 2)$},\break ${\cal L}_{\infty} = q/5$.
For deterministic allocation ($\rho\rightarrow\infty)$, the design
will ultimately be balanced (given reasonable regularity conditions on
the explanatory variables) and ${\cal L}_{\infty} = 0$. Simulation
results on the distribution of $L_n$ for other rules are presented by
\citet{aca2003a}.

The two extreme rules are random allocation and deterministic design
construction. In the completely randomized rule allocation is made independently
of any history so that the probability of allocating treatment $i$ is
$\pi_R(j) = 1/2$. For this rule
${\cal B}_\infty=0$ and ${\cal L}_\infty=q.$

In deterministic allocation the treatment with the larger value of
$d(j,n,z_{n+1})$ \eqref{5defd}
is always allocated, that is,
$ \pi_D([1]) = 1,$
where [1] is the treatment with the larger value of
$d(\cdot)$. The allocation can always be guessed so that ${\cal B}_\infty
=1$ and ${\cal L}_\infty=0.$ All other rules have intermediate values
of these two properties.

\textit{Rule~A\textup{:} Atkinson's rule.}
The remaining rules make direct use of the derivative function \eqref
{5defd}. With covariates, Atkinson's original suggestion, which is the
generalization of Smith's Rule~S \eqref{ruleS} with $\rho=2$, is
%
\begin{equation}
\pi_A(j|x_{n+1}) = \frac{d(j,n,z_{n+1})} {\sum_{s=1}^t
d(s,n,z_{n+1})}. \label{5aca5a}
\end{equation}

\textit{Rule~B\textup{:} Bayesian rule.}
Likewise, the extension of the Bayesian procedure of Section~\ref{bayessec} leads to
the rule
%
\begin{equation}
\pi_B(j|x_{n+1}) = \frac{ \{1 + d(j,n,z_{n+1})\}^{1/\gamma} } {
\sum_{s=1}^t \{1 + d(s,n,z_{n+1})\}^{1/\gamma} }. \label{5b9a}
\end{equation}
The presence of the parameter $\gamma$ is a reminder that
(\ref{5b9a}) defines a family of rules.

\begin{figure*}[b]

\includegraphics{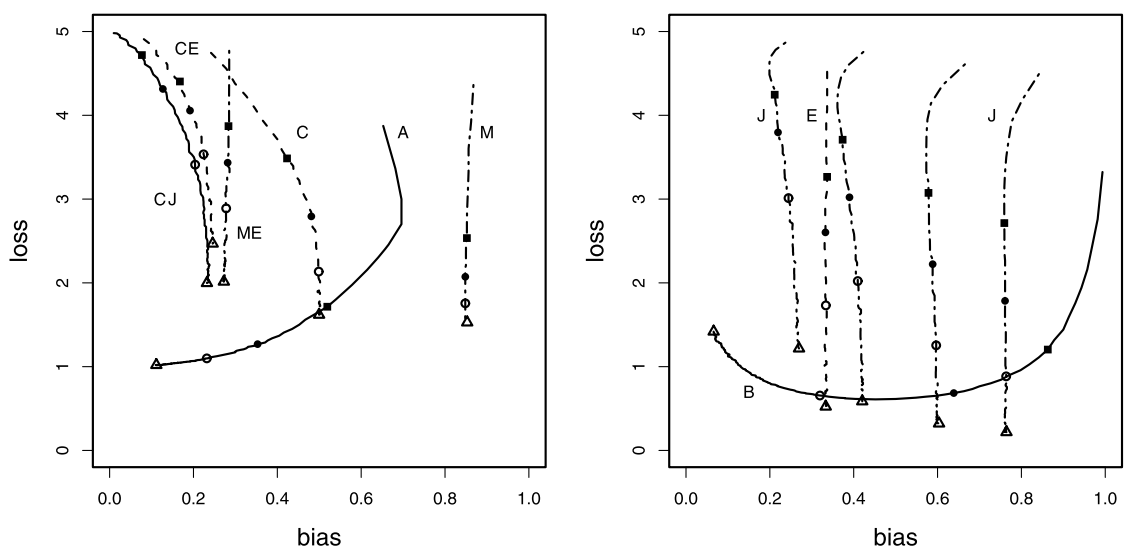}

\caption{Admissibility for normal covariates, $q = 5$. Simulated values
of loss and smoothed bias for twelve rules. Left-hand panel, curved
lines from the left---CJ(3), CE($2/3$), C and A; vertical dotted and
dashed lines---ME($2/3$) and M. Right-hand panel, vertical dotted and
dashed lines from the left---J(0.25), J(0.5), J(1) and J(2); vertical
dashed line---E($2/3$); curved line---B(0.01). Symbols: $\blacksquare\ n
= 15; \bullet\ n = 25$; $\circ\ n = 50$; $\Delta\ n = 200$. 100,000
simulations.}\label{coinf20}
\end{figure*}

\textit{Rule~E\textup{:} Generalized Efron biased-coin.}
Let [1] again be the treatment with the higher value of
$d(j,n,z_{n+1})$, the analogue of the underrepresented treatment in
Section~\ref{efrsec}. The probability of allocating this treatment is~$p$.

\textit{Rule~J\textup{:} An extension of ABCD to a model with covariates.}
To develop an analogue to Rule~J of Section~\ref{abcdblsec} requires a
relationship between the difference $D_n$ in \eqref{baldirule} and the
values of the $d(j,n,z_{n+1})$ in \eqref{5defd}.

In the absence of covariates,
%
\begin{equation}
d(1,n) = n_2/(n n_1), \label{dnocovar}
\end{equation}
with the complementary expression for $d(2,n)$. It is then
straightforward that
%
\begin{equation}
D_n = \frac{2 - n \{d(1,n) + d(2,n)\}}{d(1,n) - d(2,n)}. \label{pseudodiff}
\end{equation}
For models with covariates we calculate $D_n(z_{n+1})$ from \eqref
{pseudodiff} as a function of the $d(j,n,z_{n+1})$ and substitute for
$x$ in \eqref{baldirule}. This provides a family of rules depending on
the parameter $a$.

The difference of derivatives in the denominator of~\eqref{pseudodiff}
did not cause numerical problems in the simulations of Section~\ref{normalsec}. With discrete covariates exact balance is possible when,
from \eqref{dnocovar}, it follows that $d(1,n,z_{n+1})= d(2,n,z_{n+1})
=1/n$. The probability of assigning either treatment is then one half.
With continuous random covariates exact balance is impossible. Close to
balance, $d(1,n,z_{n+1}) \cong1/n + \varepsilon$ and $d(2,n,z_{n+1})\cong
1/n - \varepsilon\ (\varepsilon\gtrless0)$, so that $D_n(z_{n+1})
\cong0$. The probability of assigning either treatment is close to one half.

\section{Admissibility with Normal Covariates}
\label{normalsec}

The comparisons of these rules are again based on 100,000 simulations,
now with four standard normal covariates $(q = 5)$, dichotomized about
their means for rules that require discretized variables. The
regression model \eqref{5aca6} is used in the analysis of data from all
rules. Simulation results for a few rules when the covariates are
binary are given by \citet{sennval2010} and \citet{aca2012}. The
discussion here is mainly in terms of admissibility; the two panels of
Figure~\ref{coinf20} are to be compared with those of Figure~\ref{coinf12}. Additionally, values of loss and bias for $n = 50$ and 200
are given in Table~\ref{tab10}.

\begin{table}
\caption{Normal covariates, $q = 5$. Average values of loss and bias,
$\bar{B}_n$ and $\bar{L}_n$, for the twelve rules of Figure~\protect\ref
{coinf20} from 100,000 simulations for $n = 50$ and 200}\label{tab10}
\begin{tabular*}{\columnwidth}{@{\extracolsep{\fill}}lcccc@{}}
\hline
\textbf{Rule} & $\bolds{\bar{L}_{50}}$ &
$\bolds{\bar{L}_{200}}$ & $\bolds{\bar{B}_{50}}$ & \multicolumn{1}{c@{}}{$\bolds{\bar
{B}_{200}}$} \\
\hline
M & 1.7559 & 1.5275 & 0.8512 & 0.8534 \\
ME & 2.8892 & 2.0141 & 0.2799 & 0.2724 \\
C & 2.1346 & 1.6193 & 0.5035 & 0.4996 \\
CE & 3.5343 & 2.4683 & 0.2199 & 0.2464 \\
CJ(3)& 3.4106 & 1.9977 & 0.1983 & 0.2321 \\
A & 1.0985 & 1.0194 & 0.2318 & 0.1114 \\[3pt]
J(2) & 0.8845 & 0.2182 & 0.7628 & 0.7644 \\
J(1) & 1.2544 & 0.3210 & 0.5985 & 0.5967 \\
J(0.5) & 2.0214 & 0.5856 & 0.4127 & 0.4204 \\
J(0.25) & 3.0118 & 1.2165 & 0.2444 & 0.2706 \\
E & 1.7309 & 0.5229 & 0.3293 & 0.3352 \\
B & 0.6555 & 1.4183 & 0.3196 & 0.0660 \\
\hline
\end{tabular*}
\end{table}

The left-hand panel of Figure~\ref{coinf20} provides a comparison of
the more traditional rules. The discussion is from the right-hand side
of the panel, which corresponds to reading down in the table.

The unrandomized version of minimization,\break  Rule~M, has a bias of around
0.85 and a loss of 1.5275 when $n = 200$. Most of the change in the
properties of this rule, and of the adjacently plotted Rule~A, has
happened before $n = 15$, the first symbol in the plot. It is clear
that Rule~A has lower loss and lower bias than unrandomized
minimization, which is inadmissible. The other nonrandomized Rule C
follows a rather different trajectory. Initially, all cells are empty
and allocation is at random so the bias is small. However, with $q-1 =
4$ the values of the differences $D_{\iota}$ rapidly become zero or one
and the bias tends to 0.5. For most $n$ in the range 15--200, Rule~A
has lower loss and bias than Rule~C.

The randomized versions of these two rules have a similar structure of
loss and bias with $n$, but with slightly higher loss and appreciably
lower bias. For Rule~ME, with $p = 2/3$, bias is constant at a little
less than 0.3, whereas Rule~CE has a bias that tends to 0.25. Rule~CJ,
with $a = 3$, is similar in behaviour to CE initially with higher loss
when $n$ is small. But, by the time $n = 25$, CJ has both lower loss
and lower bias than CE, which is thus inadmissible, paralleling the
result in the right-hand panel of Figure~\ref{coinf12}.

The rule with lowest loss in the left-hand panel of the figure, except
for small $n$, is Rule~A. The other rules have a loss which slowly
decreases to zero. However, the balance in these rules is over
discretized values, so balance is achieved more slowly than for
deterministic allocation (Rule~D) which is based on the actual values
of the covariates. The comparisons of Rules~M, ME and D in \citet
{aca2012} show how much faster the loss of Rule~M goes to zero for
Bernoulli explanatory variables. Results in \citet{aca2002} indicate
that, apart from the doubling of loss, most rules behave similarly when
$q = 5$ and 10.

The properties of the new Rule~J of Section~\ref{covarsec} are shown in
the right-hand panel of Figure~\ref{coinf20} and in the lower part of
Table~\ref{tab10}. The right-hand curve is for $a = 2$ and the
left-hand curve for $a = 0.25$. All rules have a loss that decreases
steadily with $n$ and a bias that is virtually constant once $n = 15$.
Rule~E with $p= 2/3$ is similar to Rule~J with $a = 0.5$, but with
smaller bias and a loss which is also smaller, although the values have
become increasingly close as $n$ becomes larger. Rules~J and E both
provide a wide range of rules with constant bias and decreasing loss,
as their parameters are varied. Rule~B, also shown on the plot for
$\gamma= 0.01$, behaves very differently, but similarly to its
behaviour without covariates in Figure~\ref{coinf12}.

\section{Extensions and Conclusions}
\label{concsec}

The effect of randomization is slightly to reduce the effective sample
size by the loss and thus slightly to reduce power. \citet
{shaozhong2010} argue that, if covariates are used in randomization,
they should, as here, be included in the analysis. In discussing their
power comparisons they conclude (Section~5) that Rule~E leads to a
slightly more powerful test than that from simple randomization, a
conclusion in line with the difference in loss between the two rules.
\citet{huros2006}, Chapter~6, discuss the effect of the randomness of
loss on the distribution of power.

The majority of the randomization methods described in this paper were
developed for the comparison of two treatments and this continues to be
a major topic of research, for example, \citet{heritier2005}, \citet
{gwise2008} and \citet{linsu2012}. It is, however, straightforward
to extend Efron's rule to $t$ treatments. In the absence of covariates,
the treatments are ordered from most allocated to least allocated. The
probabilities of allocation should increase with this order. If there
are no ties, the sequence of allocation probabilities
%
\begin{equation}
\pi_E\bigl([j]\bigr) = \frac{2(t+1-j)}{t(t+1)} \label{genefr}
\end{equation}
reduces to Rule~E of Section~\ref{efrsec} when $t = 2$. Ties affect
this rule by causing
the probabilities to be averaged over the sets of tied treatments. With
covariates the treatment with the highest value of $d(j,n,z_{n+1})$
should have the highest probability of allocation. Ordering the
treatments according to the values of $d(j,n,z_{n+1})$ from smallest to
largest and applying \eqref{genefr} leads to the appropriate
generalization of Rule~E in Section~\ref{covarsec}. Rule~A with or
without covariates extends straightforwardly to any number of
treatments and is given in this form in Section~\ref{covarsec}.

\citeauthor{rls84b} (\citeyear{rls84b}) in his Section~9 formulates an allocation procedure for $t$
treatments. Let $D_{n,i}$ be the difference between the number of
patients allocated to treatment $i$ and the equal allocation target
number $n/t$. Then he shows that, asymptotically,
%
\begin{equation}
\operatorname{Var} D_{n,i} \cong n(t-1)/\bigl\{t^2(1+2\rho)\bigr\}.
\label{ttreatvar}
\end{equation}

\citet{acaatanu2004a} extend Rule~A to unequal allocation targets for
two treatments by use of a vector of target weights $p$ which occur
both in the information matrix for \eqref{5aca6} and as weights $p_j$
in \eqref{5aca5a}. The details for Rule~B are in \citet
{acaatanu2004b} with multi-treatment designs in \citet{aca2004b}. In
all cases the assumption is of additive errors of constant variance.

\citet{gwise2008} apply D- and D$_{{A}}$-optimum
designs to two-treatment heteroscedastic models without covariates;
\citet{gwise2011} extend the D-optimum calculations to several
models, again without covariates. Both papers give expressions for the
asymptotically normal distribution of $n^{-1/2}N_i$. For homoscedastic
models these results follow from the variance of $D_{n,i}$ in \eqref
{ttreatvar} and the relationship of this variance to that of $N_i$.
Then, in an extension of \eqref{dibnni} to $t$ treatments,
%
\begin{equation}
n^{-1/2} N_i \sim{\cal N}\bigl[1/t, (t-1)/\bigl
\{t^2(1+2\rho)\bigr\}\bigr]. \label{dibnnit}
\end{equation}
See Remark 3 of \citet{gwise2011} for some numbers. These papers do
not consider models with covariates. However, the general results of
\citet{baldimz2011} on the structure of the information matrix
suggest that the distribution of the $N_i$ is asymptotically
independent of the presence of covariates. Thus, \eqref{dibnnit} can be
expected to hold for homoscedastic models, independently of the value
of $q$. This assertion is supported by the unpublished simulation
results of O. Sverdlov.

A great advantage of rules such as A and B that are derived from
optimum experimental design is that they can readily be applied to a
wide variety of models. \citet{rossverd2008} compare the binomial\vadjust{\goodbreak}
version of Rule~A with balanced and \mbox{unbalanced} allocation rules,
including those with an ethical component to reduce the number of
patients receiving the inferior treatment. Unlike the earlier
comparisons of \citet{beggkalish84}, which were hampered by
computational inadequacies, \citet{rossverd2008} used a full
covariate adjusted response adaptive (CARA) scheme in which the
parameter estimates of the nonlinear models, which appear in the design
criterion, were updated before each allocation. Because choice of
treatment allocation depends on previous responses through the
parameter estimates, the observations are no longer independent. A
theory of inference for CARA designs is developed by \citet
{ZhangHu2007}. In their survey of adaptive randomization, \citet
{rossverdhu2012} are optimistic that standard inferential methods
may be used for inference in these trials, provided sample sizes are
large enough. Simulation studies can provide this reassurance.

One main purpose of this paper is to emphasize the importance of using
a measure of selection bias, as well as loss (or some other measure of
balance), in the comparison of biased-coin designs. Unfortunately,
there is a marked tendency in the literature to compare rules by
focusing on balance or loss. For example, the claim by \citeauthor
{mcent2003} for the superiority of minimization over Rule~A is based
solely on loss, ignoring randomization. \citet{sennval2010} argue that
the comparison suggested by \citet{mcent2003} is therefore potentially
misleading. However, their investigation of rules when the covariates
are binary likewise excludes any measure of bias. Similarly, the power
\mbox{comparisons} of \citet{baldi2008}, which prove the excellent properties
for power of Rule~J without covariates, do not consider bias. Most
recently, \citet{huhu2012} introduce a rule with weighted balance
over strata and covariates. However, interest is solely in analysis of
the Markov chain of the $D_n$; selection bias is not considered. It is,
however, self-evident that, if selection bias is not an issue,
deterministic construction of optimum designs, Rule~D, will provide the
lowest loss out of all myopic rules considering one treatment
allocation at a time.

There remains the choice of randomizing rule. \citet{rls84b}, Section~4,
recommends that the design should become increasingly
random as $n \rightarrow\infty$, a property of Rule~B. The rule starts
by forcing balance, which will be important if the trial stops when $n$
is small. The rate at which the allocation becomes more random
depends\vadjust{\goodbreak}
on the parameter $\gamma$; Figure~8 of \citet{aca2002} shows
admissibility curves for Rule B for six values of $\gamma$.

Of course, as the design becomes more random, loss increases and so it
might be suspected that the efficiency
\eqref{nocoveffic} would decrease. However, for Rule~B, $L_n$ increases
up to the limit $q$ and is divided by $n$ in~\eqref{nocoveffic}.
Figure~4 of \citet{aca2002} shows how the efficiency increases to one
with $n$, at a rate depending on the value of $\gamma$. For the results
in Table~\ref{tab10} for Rule~B with $\gamma= 0.01, \bar{L}_{200} =
1.4183$, so that the efficiency is 99.29\%. With an average bias at
this point of 0.0660, the rule virtually has the bias of random
allocation. Of course, an adjustable randomization rule is
administratively more complicated than one with a constant probability,
although hardly more so than any other rule that takes account of the
covariates of each patient and certainly less so than a response
adaptive rule. \citet{sverdbill2013} are hopeful that developments in
computing and information science will enable routine use of
randomization rules more complicated than Rule~B.

\section*{Acknowledgements} I am grateful to the referees for their
positive comments which led to the extension of this paper to rules
including covariates. I have also enjoyed and profited from
conversations with Dr O. (Alex) Sverdlov.


%




\begin{thebibliography}{48}



\bibitem[\protect\citeauthoryear{Atkinson}{1982}]{aca82}
\begin{barticle}[mr]
\bauthor{\bsnm{Atkinson},~\bfnm{A.~C.}\binits{A.~C.}}
(\byear{1982}).
\btitle{Optimum biased coin designs for sequential clinical trials with
prognostic factors}.
\bjournal{Biometrika}
\bvolume{69}
\bpages{61--67}.
\bid{doi={10.1093/biomet/69.1.61}, issn={0006-3444}, mr={0655670}}
\bptok{imsref}%
\end{barticle}
\endbibitem

\bibitem[\protect\citeauthoryear{Atkinson}{2002}]{aca2002}
\begin{barticle}[mr]
\bauthor{\bsnm{Atkinson},~\bfnm{Anthony~C.}\binits{A.~C.}}
(\byear{2002}).
\btitle{The comparison of designs for sequential clinical trials with covariate
information}.
\bjournal{J. Roy. Statist. Soc. Ser. A}
\bvolume{165}
\bpages{349--373}.
\bid{doi={10.1111/1467-985X.00564}, issn={0964-1998}, mr={1904822}}
\bptok{imsref}%
\end{barticle}
\endbibitem

\bibitem[\protect\citeauthoryear{Atkinson}{2003}]{aca2003a}
\begin{barticle}[pbm]
\bauthor{\bsnm{Atkinson},~\bfnm{Anthony~C.}\binits{A.~C.}}
(\byear{2003}).
\btitle{The distribution of loss in two-treatment biased-coin designs}.
\bjournal{Biostatistics}
\bvolume{4}
\bpages{179--193}.
\bid{doi={10.1093/biostatistics/4.2.179}, issn={1465-4644}, pii={4/2/179},
pmid={12925515}}
\bptok{imsref}%
\end{barticle}
\endbibitem

\bibitem[\protect\citeauthoryear{Atkinson}{2004}]{aca2004b}
\begin{barticle}[mr]
\bauthor{\bsnm{Atkinson},~\bfnm{Anthony~C.}\binits{A.~C.}}
(\byear{2004}).
\btitle{Adaptive biased-coin designs for clinical trials with several
treatments}.
\bjournal{Discuss. Math. Probab. Stat.}
\bvolume{24}
\bpages{85--108}.
\bid{issn={1509-9423}, mr={2118925}}
\bptok{imsref}%
\end{barticle}
\endbibitem

\bibitem[\protect\citeauthoryear{Atkinson}{2012}]{aca2012}
\begin{barticle}[mr]
\bauthor{\bsnm{Atkinson},~\bfnm{Anthony~C.}\binits{A.~C.}}
(\byear{2012}).
\btitle{Bias and loss: The two sides of a biased coin}.
\bjournal{Stat. Med.}
\bvolume{31}
\bpages{3494--3503}.
\bid{doi={10.1002/sim.5416}, issn={0277-6715}, mr={3041826}}
\bptok{imsref}%
\end{barticle}
\endbibitem

\bibitem[\protect\citeauthoryear{Atkinson and Biswas}{2005a}]{acaatanu2004b}
\begin{barticle}[mr]
\bauthor{\bsnm{Atkinson},~\bfnm{Anthony~C.}\binits{A.~C.}} \AND
\bauthor{\bsnm{Biswas},~\bfnm{Atanu}\binits{A.}}
(\byear{2005}a).
\btitle{Bayesian adaptive biased-coin designs for clinical trials with normal
responses}.
\bjournal{Biometrics}
\bvolume{61}
\bpages{118--125}.
\bid{doi={10.1111/j.0006-341X.2005.031002.x}, issn={0006-341X}, mr={2135851}}
\bptok{imsref}%
\end{barticle}
\endbibitem

\bibitem[\protect\citeauthoryear{Atkinson and Biswas}{2005b}]{acaatanu2004a}
\begin{barticle}[mr]
\bauthor{\bsnm{Atkinson},~\bfnm{A.~C.}\binits{A.~C.}} \AND
\bauthor{\bsnm{Biswas},~\bfnm{A.}\binits{A.}}
(\byear{2005}b).
\btitle{Adaptive biased-coin designs for skewing the allocation proportion in
clinical trials with normal responses}.
\bjournal{Stat. Med.}
\bvolume{24}
\bpages{2477--2492}.
\bid{doi={10.1002/sim.2124}, issn={0277-6715}, mr={2112377}}
\bptok{imsref}%
\end{barticle}
\endbibitem

\bibitem[\protect\citeauthoryear{Atkinson and Biswas}{2014}]{acaatanu2014a}
\begin{bbook}[auto:STB|2013/12/09|07:59:19]
\bauthor{\bsnm{Atkinson},~\bfnm{A.~C.}\binits{A.~C.}} \AND
\bauthor{\bsnm{Biswas},~\bfnm{A.}\binits{A.}}
(\byear{2014}).
\btitle{Randomised Response-Adaptive Designs in Clinical Trials}.
\bpublisher{Chapman \& Hall/CRC Press}, \blocation{Boca Raton}.
\bptok{imsref}%
\end{bbook}
\endbibitem

\bibitem[\protect\citeauthoryear{Atkinson, Donev and Tobias}{2007}]{ADT2007}
\begin{bbook}[mr]
\bauthor{\bsnm{Atkinson},~\bfnm{A.~C.}\binits{A.~C.}},
\bauthor{\bsnm{Donev},~\bfnm{A.~N.}\binits{A.~N.}} \AND
\bauthor{\bsnm{Tobias},~\bfnm{R.~D.}\binits{R.~D.}}
(\byear{2007}).
\btitle{Optimum Experimental Designs, with {SAS}}.
\bseries{Oxford Statistical Science Series}
\bvolume{34}.
\bpublisher{Oxford Univ. Press}, \blocation{Oxford}.
\bid{mr={2323647}}
\bptok{imsref}%
\end{bbook}
\endbibitem

\bibitem[\protect\citeauthoryear{Bailey and Nelson}{2003}]{rabpn2003}
\begin{barticle}[mr]
\bauthor{\bsnm{Bailey},~\bfnm{R.~A.}\binits{R.~A.}} \AND
\bauthor{\bsnm{Nelson},~\bfnm{P.~R.}\binits{P.~R.}}
(\byear{2003}).
\btitle{Hadamard randomization: A valid restriction of random permuted blocks}.
\bjournal{Biom. J.}
\bvolume{45}
\bpages{554--560}.
\bid{doi={10.1002/bimj.200390032}, issn={0323-3847}, mr={1998135}}
\bptok{imsref}%
\end{barticle}
\endbibitem

\bibitem[\protect\citeauthoryear{Baldi Antognini}{2008}]{baldi2008}
\begin{barticle}[mr]
\bauthor{\bsnm{Baldi Antognini},~\bfnm{Alessandro}\binits{A.}}
(\byear{2008}).
\btitle{A theoretical analysis of the power of biased coin designs}.
\bjournal{J. Statist. Plann. Inference}
\bvolume{138}
\bpages{1792--1798}.
\bid{doi={10.1016/j.jspi.2007.06.033}, issn={0378-3758}, mr={2400479}}
\bptok{imsref}%
\end{barticle}
\endbibitem


\bibitem[\protect\citeauthoryear{Baldi~Antognini and
Giovagnoli}{2004}]{baldis2004}
\begin{barticle}[mr]
\bauthor{\bsnm{Baldi~Antognini},~\bfnm{Alessandro}\binits{A.}} \AND
\bauthor{\bsnm{Giovagnoli},~\bfnm{Alessandra}\binits{A.}}
(\byear{2004}).
\btitle{A new ``biased coin design'' for the sequential allocation of two
treatments}.
\bjournal{J. R. Stat. Soc. Ser. C Appl. Stat.}
\bvolume{53}
\bpages{651--664}.
\bid{doi={10.1111/j.1467-9876.2004.00436.x}, issn={0035-9254}, mr={2087777}}
\bptok{imsref}%
\end{barticle}
\endbibitem

\bibitem[\protect\citeauthoryear{Baldi~Antognini and
Zagoraiou}{2011}]{baldimz2011}
\begin{barticle}[mr]
\bauthor{\bsnm{Baldi~Antognini},~\bfnm{A.}\binits{A.}} \AND
\bauthor{\bsnm{Zagoraiou},~\bfnm{M.}\binits{M.}}
(\byear{2011}).
\btitle{The covariate-adaptive biased coin design for balancing clinical trials
in the presence of prognostic factors}.
\bjournal{Biometrika}
\bvolume{98}
\bpages{519--535}.
\bid{doi={10.1093/biomet/asr021}, issn={0006-3444}, mr={2836404}}
\bptok{imsref}%
\end{barticle}
\endbibitem

\bibitem[\protect\citeauthoryear{Ball, Smith and Verdinelli}{1993}]{ball93}
\begin{barticle}[mr]
\bauthor{\bsnm{Ball},~\bfnm{F.~G.}\binits{F.~G.}},
\bauthor{\bsnm{Smith},~\bfnm{A.~F.~M.}\binits{A.~F.~M.}} \AND
\bauthor{\bsnm{Verdinelli},~\bfnm{I.}\binits{I.}}
(\byear{1993}).
\btitle{Biased coin designs with a {B}ayesian bias}.
\bjournal{J. Statist. Plann. Inference}
\bvolume{34}
\bpages{403--421}.
\bid{doi={10.1016/0378-3758(93)90148-Y}, issn={0378-3758}, mr={1210443}}
\bptok{imsref}%
\end{barticle}
\endbibitem

\bibitem[\protect\citeauthoryear{Begg and Kalish}{1984}]{beggkalish84}
\begin{barticle}[pbm]
\bauthor{\bsnm{Begg},~\bfnm{C.~B.}\binits{C.~B.}} \AND
\bauthor{\bsnm{Kalish},~\bfnm{L.~A.}\binits{L.~A.}}
(\byear{1984}).
\btitle{Treatment allocation for nonlinear models in clinical trials: The
logistic model}.
\bjournal{Biometrics}
\bvolume{40}
\bpages{409--420}.
\bid{issn={0006-341X}, pmid={6487725}}
\bptok{imsref}%
\end{barticle}
\endbibitem

\bibitem[\protect\citeauthoryear{Berger}{2005}]{berg2005}
\begin{bbook}[auto:STB|2013/12/09|07:59:19]
\bauthor{\bsnm{Berger},~\bfnm{V.~W.}\binits{V.~W.}}
(\byear{2005}).
\btitle{Selection Bias and Covariate Imbalances in Clinical Trials}.
\bpublisher{Wiley}, \blocation{New York}.
\bptok{imsref}%
\end{bbook}
\endbibitem


\bibitem[\protect\citeauthoryear{Biswas and Bhattacharya}{2011}]{biswasbhattacharya11}
\begin{bincollection}[auto]
\bauthor{\bsnm{Biswas},~\bfnm{A.}\binits{A.}} \AND
\bauthor{\bsnm{Bhattacharya},~\bfnm{R.}\binits{R.}}
(\byear{2011}).
\btitle{Treatment adaptive allocations in randomized clinical trials: An
 overview}.
In \bbooktitle{Handbook of Adaptive Designs
 in Pharmaceutical and Clinical Development}
(\beditor{\bfnm{A.}\binits{A.}~\bsnm{Pong}} \AND
\beditor{\bfnm{S.-C.}\binits{S.-C.}~\bsnm{Chow}}, eds.)
\bpages{17:1--17:19}.
\bpublisher{Chapman \& Hall/CRC Press}, \blocation{Boca Raton, FL}.
\bptok{imsref}%
\end{bincollection}
\endbibitem


\bibitem[\protect\citeauthoryear{Blackwell and Hodges}{1957}]{bh57}
\begin{barticle}[mr]
\bauthor{\bsnm{Blackwell},~\bfnm{David}\binits{D.}} \AND
\bauthor{\bsnm{Hodges},~\bfnm{J.~L.}\binits{J.~L.} \bsuffix{Jr.}}
(\byear{1957}).
\btitle{Design for the control of selection bias}.
\bjournal{Ann. Math. Statist.}
\bvolume{28}
\bpages{449--460}.
\bid{issn={0003-4851}, mr={0088849}}
\bptok{imsref}%
\end{barticle}
\endbibitem

\bibitem[\protect\citeauthoryear{Burman}{1996}]{bur96}
\begin{bmisc}[auto:STB|2013/12/09|07:59:19]
\bauthor{\bsnm{Burman},~\bfnm{C.~F.}\binits{C.~F.}}
(\byear{1996}).
\bhowpublished{\textit{On Sequential Treatment Allocations in Clinical Trials}.
Dept. Mathematics, G\"oteborg}.
\bptok{imsref}%
\end{bmisc}
\endbibitem

\bibitem[\protect\citeauthoryear{Chen}{1999}]{ypchen99}
\begin{barticle}[mr]
\bauthor{\bsnm{Chen},~\bfnm{Yung-Pin}\binits{Y.-P.}}
(\byear{1999}).
\btitle{Biased coin design with imbalance tolerance}.
\bjournal{Comm. Statist. Stochastic Models}
\bvolume{15}
\bpages{953--975}.
\bid{doi={10.1080/15326349908807570}, issn={0882-0287}, mr={1721241}}
\bptok{imsref}%
\end{barticle}
\endbibitem

\bibitem[\protect\citeauthoryear{Cox}{1951}]{cox51}
\begin{barticle}[mr]
\bauthor{\bsnm{Cox},~\bfnm{D.~R.}\binits{D.~R.}}
(\byear{1951}).
\btitle{Some systematic experimental designs}.
\bjournal{Biometrika}
\bvolume{38}
\bpages{312--323}.
\bid{issn={0006-3444}, mr={0046013}}
\bptok{imsref}%
\end{barticle}
\endbibitem

\bibitem[\protect\citeauthoryear{Efron}{1971}]{efr71}
\begin{barticle}[mr]
\bauthor{\bsnm{Efron},~\bfnm{Bradley}\binits{B.}}
(\byear{1971}).
\btitle{Forcing a sequential experiment to be balanced}.
\bjournal{Biometrika}
\bvolume{58}
\bpages{403--417}.
\bid{issn={0006-3444}, mr={0312660}}
\bptok{imsref}%
\end{barticle}
\endbibitem

\bibitem[\protect\citeauthoryear{Efron}{1980}]{brad80}
\begin{bincollection}[auto:STB|2013/12/09|07:59:19]
\bauthor{\bsnm{Efron},~\bfnm{B.}\binits{B.}}
(\byear{1980}).
\btitle{Randomizing and balancing a complicated sequential experiment}.
In \bbooktitle{Biostatistics Casebook}
(\beditor{\bfnm{R.~J.}\binits{R.~J.}~\bsnm{Miller}},
\beditor{\bfnm{B.}\binits{B.}~\bsnm{Efron}},
\beditor{\bfnm{B.~W.}\binits{B.~W.}~\bsnm{Brown}} \AND
\beditor{\bfnm{L.~E.}\binits{L.~E.}~\bsnm{Moses}}, eds.)
\bpages{19--30}.
\bpublisher{Wiley}, \blocation{New York}.
\bptok{imsref}%
\end{bincollection}
\endbibitem

\bibitem[\protect\citeauthoryear{Gwise, Hu and Hu}{2008}]{gwise2008}
\begin{barticle}[mr]
\bauthor{\bsnm{Gwise},~\bfnm{Thomas~E.}\binits{T.~E.}},
\bauthor{\bsnm{Hu},~\bfnm{Jianhua}\binits{J.}} \AND
\bauthor{\bsnm{Hu},~\bfnm{Feifang}\binits{F.}}
(\byear{2008}).
\btitle{Optimal biased coins for two-arm clinical trials}.
\bjournal{Stat. Interface}
\bvolume{1}
\bpages{125--135}.
\bid{doi={10.4310/SII.2008.v1.n1.a11}, issn={1938-7989}, mr={2425350}}
\bptok{imsref}%
\end{barticle}
\endbibitem

\bibitem[\protect\citeauthoryear{Gwise, Zhou and Hu}{2011}]{gwise2011}
\begin{barticle}[mr]
\bauthor{\bsnm{Gwise},~\bfnm{Thomas~E.}\binits{T.~E.}},
\bauthor{\bsnm{Zhou},~\bfnm{Jianhui}\binits{J.}} \AND
\bauthor{\bsnm{Hu},~\bfnm{Feifang}\binits{F.}}
(\byear{2011}).
\btitle{An optimal response adaptive biased coin design with {$k$}
heteroscedastic treatments}.
\bjournal{J. Statist. Plann. Inference}
\bvolume{141}
\bpages{235--242}.
\bid{doi={10.1016/j.jspi.2010.06.013}, issn={0378-3758}, mr={2719490}}
\bptok{imsref}%
\end{barticle}
\endbibitem

\bibitem[\protect\citeauthoryear{Heritier, Gebski and
Pillai}{2005}]{heritier2005}
\begin{barticle}[mr]
\bauthor{\bsnm{Heritier},~\bfnm{Stephane}\binits{S.}},
\bauthor{\bsnm{Gebski},~\bfnm{Val}\binits{V.}} \AND
\bauthor{\bsnm{Pillai},~\bfnm{Avinesh}\binits{A.}}
(\byear{2005}).
\btitle{Dynamic balancing randomization in controlled clinical trials}.
\bjournal{Stat. Med.}
\bvolume{24}
\bpages{3729--3741}.
\bid{doi={10.1002/sim.2421}, issn={0277-6715}, mr={2221964}}
\bptok{imsref}%
\end{barticle}
\endbibitem

\bibitem[\protect\citeauthoryear{Hu and Hu}{2012}]{huhu2012}
\begin{barticle}[mr]
\bauthor{\bsnm{Hu},~\bfnm{Yanqing}\binits{Y.}} \AND
\bauthor{\bsnm{Hu},~\bfnm{Feifang}\binits{F.}}
(\byear{2012}).
\btitle{Asymptotic properties of covariate-adaptive randomization}.
\bjournal{Ann. Statist.}
\bvolume{40}
\bpages{1794--1815}.
\bid{doi={10.1214/12-AOS983}, issn={0090-5364}, mr={3015044}}
\bptok{imsref}%
\end{barticle}
\endbibitem

\bibitem[\protect\citeauthoryear{Hu and Rosenberger}{2006}]{huros2006}
\begin{bbook}[mr]
\bauthor{\bsnm{Hu},~\bfnm{Feifang}\binits{F.}} \AND
\bauthor{\bsnm{Rosenberger},~\bfnm{William~F.}\binits{W.~F.}}
(\byear{2006}).
\btitle{The Theory of Response-Adaptive Randomization in Clinical Trials}.
\bpublisher{Wiley}, \blocation{Hoboken, NJ}.
\bid{doi={10.1002/047005588X}, mr={2245329}}
\bptok{imsref}%
\end{bbook}
\endbibitem

\bibitem[\protect\citeauthoryear{Lagakos and Pocock}{1984}]{lagakos1984}
\begin{bincollection}[auto:STB|2013/12/09|07:59:19]
\bauthor{\bsnm{Lagakos},~\bfnm{S.~W.}\binits{S.~W.}} \AND
\bauthor{\bsnm{Pocock},~\bfnm{S.~J.}\binits{S.~J.}}
(\byear{1984}).
\btitle{Randomization and stratification in cancer clinical trials: An
international survey}.
In \bbooktitle{Cancer Clinical Trials: Methods and Practice}
(\beditor{\bfnm{M.~E.}\binits{M.~E.}~\bsnm{Buyse}},
\beditor{\bfnm{M.~J.}\binits{M.~J.}~\bsnm{Staquet}} \AND
\beditor{\bfnm{R.~J.}\binits{R.~J.}~\bsnm{Sylvester}}, eds.).
\bpublisher{Oxford Univ. Press}, \blocation{Oxford}.
\bptok{imsref}%
\end{bincollection}
\endbibitem

\bibitem[\protect\citeauthoryear{Lin and Su}{2012}]{linsu2012}
\begin{barticle}[mr]
\bauthor{\bsnm{Lin},~\bfnm{Yunzhi}\binits{Y.}} \AND
\bauthor{\bsnm{Su},~\bfnm{Zheng}\binits{Z.}}
(\byear{2012}).
\btitle{Balancing continuous and categorical baseline covariates in sequential
clinical trials using the area between empirical cumulative distribution
functions}.
\bjournal{Stat. Med.}
\bvolume{31}
\bpages{1961--1971}.
\bid{doi={10.1002/sim.5363}, issn={0277-6715}, mr={2956029}}
\bptok{imsref}%
\end{barticle}
\endbibitem

\bibitem[\protect\citeauthoryear{Markaryan and
Rosenberger}{2010}]{markbill2010}
\begin{barticle}[mr]
\bauthor{\bsnm{Markaryan},~\bfnm{Tigran}\binits{T.}} \AND
\bauthor{\bsnm{Rosenberger},~\bfnm{William~F.}\binits{W.~F.}}
(\byear{2010}).
\btitle{Exact properties of {E}fron's biased coin randomization procedure}.
\bjournal{Ann. Statist.}
\bvolume{38}
\bpages{1546--1567}.
\bid{doi={10.1214/09-AOS758}, issn={0090-5364}, mr={2662351}}
\bptok{imsref}%
\end{barticle}
\endbibitem

\bibitem[\protect\citeauthoryear{McEntegart}{2003}]{mcent2003}
\begin{barticle}[auto:STB|2013/12/09|07:59:19]
\bauthor{\bsnm{McEntegart},~\bfnm{D.}\binits{D.}}
(\byear{2003}).
\btitle{The pursuit of balance using stratified and dynamic randomization
techniques: An overview}.
\bjournal{Drug Information Journal}
\bvolume{37}
\bpages{293--308}.
\bptok{imsref}%
\end{barticle}
\endbibitem

\bibitem[\protect\citeauthoryear{Pocock}{1983}]{pocock83}
\begin{bbook}[auto:STB|2013/12/09|07:59:19]
\bauthor{\bsnm{Pocock},~\bfnm{S.~J.}\binits{S.~J.}}
(\byear{1983}).
\btitle{Clinical Trials}.
\bpublisher{Wiley}, \blocation{New York}.
\bptok{imsref}%
\end{bbook}
\endbibitem

\bibitem[\protect\citeauthoryear{Pocock and Simon}{1975}]{ps75}
\begin{barticle}[pbm]
\bauthor{\bsnm{Pocock},~\bfnm{S.~J.}\binits{S.~J.}} \AND
\bauthor{\bsnm{Simon},~\bfnm{R.}\binits{R.}}
(\byear{1975}).
\btitle{Sequential treatment assignment with balancing for prognostic factors
in the controlled clinical trial}.
\bjournal{Biometrics}
\bvolume{31}
\bpages{103--\break 115}.
\bid{issn={0006-341X}, pmid={1100130}}
\bptok{imsref}%
\end{barticle}
\endbibitem




\bibitem[\protect\citeauthoryear{Proschan, Brittain and
Kammerman}{2012}]{prosch2012}
\begin{barticle}[auto:STB|2013/12/09|07:59:19]
\bauthor{\bsnm{Proschan},~\bfnm{M.}\binits{M.}},
\bauthor{\bsnm{Brittain},~\bfnm{E.}\binits{E.}} \AND
\bauthor{\bsnm{Kammerman},~\bfnm{L.}\binits{L.}}
(\byear{2012}).
\btitle{Letter to the editor}.
\bjournal{Biometrics}
\bvolume{68}
\bpages{990--991}.
\bptok{imsref}%
\end{barticle}
\endbibitem

\bibitem[\protect\citeauthoryear{Rosenberger and Lachin}{2002}]{rosl2002}
\begin{bbook}[mr]
\bauthor{\bsnm{Rosenberger},~\bfnm{William~F.}\binits{W.~F.}} \AND
\bauthor{\bsnm{Lachin},~\bfnm{John~M.}\binits{J.~M.}}
(\byear{2002}).
\btitle{Randomization in Clinical Trials: Theory and Practice}.
\bpublisher{Wiley}, \blocation{New York}.
\bid{doi={10.1002/0471722103}, mr={1914364}}
\bptok{imsref}%
\end{bbook}
\endbibitem

\bibitem[\protect\citeauthoryear{Rosenberger and Sverdlov}{2008}]{rossverd2008}
\begin{barticle}[mr]
\bauthor{\bsnm{Rosenberger},~\bfnm{William~F.}\binits{W.~F.}} \AND
\bauthor{\bsnm{Sverdlov},~\bfnm{Oleksandr}\binits{O.}}
(\byear{2008}).
\btitle{Handling covariates in the design of clinical trials}.
\bjournal{Statist. Sci.}
\bvolume{23}
\bpages{404--419}.
\bid{doi={10.1214/08-STS269}, issn={0883-4237}, mr={2483911}}
\bptok{imsref}%
\end{barticle}
\endbibitem

\bibitem[\protect\citeauthoryear{Rosenberger, Sverdlov and
Hu}{2012}]{rossverdhu2012}
\begin{barticle}[mr]
\bauthor{\bsnm{Rosenberger},~\bfnm{William~F.}\binits{W.~F.}},
\bauthor{\bsnm{Sverdlov},~\bfnm{Oleksandr}\binits{O.}} \AND
\bauthor{\bsnm{Hu},~\bfnm{Feifang}\binits{F.}}
(\byear{2012}).
\btitle{Adaptive randomization for clinical trials}.
\bjournal{J. Biopharm. Statist.}
\bvolume{22}
\bpages{719--736}.
\bid{doi={10.1080/10543406.2012.676535}, issn={1054-3406}, mr={2931067}}
\bptok{imsref}%
\end{barticle}
\endbibitem

\bibitem[\protect\citeauthoryear{Senn, Anisimov and
Fedorov}{2010}]{sennval2010}
\begin{barticle}[mr]
\bauthor{\bsnm{Senn},~\bfnm{Stephen}\binits{S.}},
\bauthor{\bsnm{Anisimov},~\bfnm{Vladimir~V.}\binits{V.~V.}} \AND
\bauthor{\bsnm{Fedorov},~\bfnm{Valerii~V.}\binits{V.~V.}}
(\byear{2010}).
\btitle{Comparisons of minimization and {A}tkinson's algorithm}.
\bjournal{Stat. Med.}
\bvolume{29}
\bpages{721--730}.
\bid{doi={10.1002/sim.3763}, issn={0277-6715}, mr={2752037}}
\bptok{imsref}%
\end{barticle}
\endbibitem

\bibitem[\protect\citeauthoryear{Shao, Yu and Zhong}{2010}]{shaozhong2010}
\begin{barticle}[mr]
\bauthor{\bsnm{Shao},~\bfnm{Jun}\binits{J.}},
\bauthor{\bsnm{Yu},~\bfnm{Xinxin}\binits{X.}} \AND
\bauthor{\bsnm{Zhong},~\bfnm{Bob}\binits{B.}}
(\byear{2010}).
\btitle{A theory for testing hypotheses under covariate-adaptive
randomization}.
\bjournal{Biometrika}
\bvolume{97}
\bpages{347--360}.
\bid{doi={10.1093/biomet/asq014}, issn={0006-3444}, mr={2650743}}
\bptok{imsref}%
\end{barticle}
\endbibitem

\bibitem[\protect\citeauthoryear{Smith}{1984a}]{rls84a}
\begin{barticle}[mr]
\bauthor{\bsnm{Smith},~\bfnm{Richard~L.}\binits{R.~L.}}
(\byear{1984}a).
\btitle{Properties of biased coin designs in sequential clinical trials}.
\bjournal{Ann. Statist.}
\bvolume{12}
\bpages{1018--1034}.
\bid{doi={10.1214/aos/1176346718}, issn={0090-5364}, mr={0751289}}
\bptok{imsref}%
\end{barticle}
\endbibitem

\bibitem[\protect\citeauthoryear{Smith}{1984b}]{rls84b}
\begin{barticle}[mr]
\bauthor{\bsnm{Smith},~\bfnm{Richard~L.}\binits{R.~L.}}
(\byear{1984}b).
\btitle{Sequential treatment allocation using biased coin designs}.
\bjournal{J. R. Stat. Soc. Ser. B Stat. Methodol.}
\bvolume{46}
\bpages{519--543}.
\bid{issn={0035-9246}, mr={0790636}}
\bptok{imsref}%
\end{barticle}
\endbibitem

\bibitem[\protect\citeauthoryear{Soares and Wu}{1983}]{soareswu83}
\begin{barticle}[mr]
\bauthor{\bsnm{Soares},~\bfnm{Jos{\'e}~F.}\binits{J.~F.}} \AND
\bauthor{\bsnm{Wu},~\bfnm{C.~F.~Jeff}\binits{C.~F.~J.}}
(\byear{1983}).
\btitle{Some restricted randomization rules in sequential designs}.
\bjournal{Comm. Statist. Theory Methods}
\bvolume{12}
\bpages{2017--2034}.
\bid{doi={10.1080/03610928308828586}, issn={0361-0926}, mr={0714209}}
\bptok{imsref}%
\end{barticle}
\endbibitem

\bibitem[\protect\citeauthoryear{Sverdlov and
Rosenberger}{2013}]{sverdbill2013}
\begin{barticle}[auto:STB|2013/12/09|07:59:19]
\bauthor{\bsnm{Sverdlov},~\bfnm{O.}\binits{O.}} \AND
\bauthor{\bsnm{Rosenberger},~\bfnm{W.~F.}\binits{W.~F.}}
(\byear{2013}).
\btitle{Randomization in clinical trials: Can we eliminate bias?}
\bjournal{Clinical Trial Perspective}
\bvolume{3}
\bpages{37--47}.
\bptok{imsref}%
\end{barticle}
\endbibitem

\bibitem[\protect\citeauthoryear{Wei}{1978}]{wei78}
\begin{barticle}[mr]
\bauthor{\bsnm{Wei},~\bfnm{L.~J.}\binits{L.~J.}}
(\byear{1978}).
\btitle{The adaptive biased coin design for sequential experiments}.
\bjournal{Ann. Statist.}
\bvolume{6}
\bpages{92--100}.
\bid{issn={0090-5364}, mr={0471205}}
\bptok{imsref}%
\end{barticle}
\endbibitem

\bibitem[\protect\citeauthoryear{Zhang et~al.}{2007}]{ZhangHu2007}
\begin{barticle}[mr]
\bauthor{\bsnm{Zhang},~\bfnm{Li-Xin}\binits{L.-X.}},
\bauthor{\bsnm{Hu},~\bfnm{Feifang}\binits{F.}},
\bauthor{\bsnm{Cheung},~\bfnm{Siu~Hung}\binits{S.~H.}} \AND
\bauthor{\bsnm{Chan},~\bfnm{Wai~Sum}\binits{W.~S.}}
(\byear{2007}).
\btitle{Asymptotic properties of covariate-adjusted response-adaptive designs}.
\bjournal{Ann. Statist.}
\bvolume{35}
\bpages{1166--1182}.
\bid{doi={10.1214/009053606000001424}, issn={0090-5364}, mr={2341702}}
\bptok{imsref}%
\end{barticle}
\endbibitem

\bibitem[\protect\citeauthoryear{Zhao et~al.}{2012}]{zhao2011}
\begin{barticle}[pbm]
\bauthor{\bsnm{Zhao},~\bfnm{Wenle}\binits{W.}},
\bauthor{\bsnm{Weng},~\bfnm{Yanqiu}\binits{Y.}},
\bauthor{\bsnm{Wu},~\bfnm{Qi}\binits{Q.}} \AND
\bauthor{\bsnm{Palesch},~\bfnm{Yuko}\binits{Y.}}
(\byear{2012}).
\btitle{Quantitative comparison of randomization designs in sequential clinical
trials based on treatment balance and allocation randomness}.
\bjournal{Pharm. Stat.}
\bvolume{11}
\bpages{39--48}.
\bid{doi={10.1002/pst.493}, issn={1539-1612}, mid={NIHMS384534},
pmcid={3399213}, pmid={21544929}}
\bptok{imsref}%
\end{barticle}
\endbibitem

\end{thebibliography}
\end{document}